\documentclass[aps,prl,superscriptaddress,twocolumn]{revtex4-2}
\usepackage{graphicx} % Include Fig.~files
\usepackage{subfigure}
\usepackage{multirow}
\usepackage{xspace}
\usepackage{fancyhdr}
\usepackage{amsmath,amssymb}
\usepackage{xcolor}

\begin{document}
\title{Ultrafast Raman probe of the photoinduced superconducting to normal state transition in the cuprate Bi$_2$Sr$_2$CaCu$_2$O$_{8+\delta}$}

\author{Laurène Gatuingt}
\email{laurene.gatuingt@u-paris.fr}
\affiliation{Université Paris Cité, Matériaux et Phénomènes Quantiques UMR CNRS 7162, Bâtiment Condorcet, 75205 Paris Cedex 13, France}
\author{Alexandr Alekhin}
\affiliation{Université Paris Cité, Matériaux et Phénomènes Quantiques UMR CNRS 7162, Bâtiment Condorcet, 75205 Paris Cedex 13, France}
\author{Niloufar Nilforoushan}
\affiliation{Université Paris Cité, Matériaux et Phénomènes Quantiques UMR CNRS 7162, Bâtiment Condorcet, 75205 Paris Cedex 13, France}
\author{Sarah Houver}
\affiliation{Université Paris Cité, Matériaux et Phénomènes Quantiques UMR CNRS 7162, Bâtiment Condorcet, 75205 Paris Cedex 13, France}
\author{Alain Sacuto}
\affiliation{Université Paris Cité, Matériaux et Phénomènes Quantiques UMR CNRS 7162, Bâtiment Condorcet, 75205 Paris Cedex 13, France}
\author{Genda Gu}
\affiliation{Brookhaven National Lab, Upton, NY, USA}
\author{Yann Gallais}
\email{yann.gallais@u-paris.fr}
\affiliation{Université Paris Cité, Matériaux et Phénomènes Quantiques UMR CNRS 7162, Bâtiment Condorcet, 75205 Paris Cedex 13, France}

\begin{abstract}
We report an ultrafast Time-Resolved Raman scattering study of the out-of-equilibrium photoinduced dynamics across the superconducting to normal state phase transition of the cuprate Bi$_2$Sr$_2$CaCu$_2$O$_{8+\delta}$. Using the polarization-resolved momentum space selectivity of Raman scattering, we track the superconducting condensate destruction and recovery dynamics with sub-picoseconds time resolution in the anti-nodal region of the Fermi surface where the superconducting gap is maximum. Leveraging ultrafast Raman thermometry, we find a significant dichotomy between the superconducting condensate and the quasiparticle temperature dynamics near the anti-nodes, which cannot be framed in terms of a single effective electron temperature. The present work demonstrates the ability of Time-Resolved Raman scattering to selectively probe out-of-equilibrium pathways of different electronic sub-degrees of freedom during a photoinduced phase transition.
\end{abstract}

\maketitle

\section{Introduction}
Ultrafast photoinduced phase transitions are a particular class of out-of-equilibrium phase transitions triggered by intense light pulses, offering the potential to transiently engineer phases of matter that are inaccessible under equilibrium conditions \cite{yonemitsu_theory_2008,de_la_torre_colloquium_2021}. These studies have been enabled by recent progress in cutting-edge experimental tools capable of monitoring materials properties at femtosecond (fs) time scales. Among photoinduced phase transitions, those involving electronic orders such as charge density waves or superconductivity have been extensively studied. These investigations have revealed distinct ultrafast dynamics of different degrees of freedom, including the lattice and electron subsystems. \cite{giannetti_ultrafast_2016,pashkin_femtosecond_2010,porer_non-thermal_2014,maklar_nonequilibrium_2021,konstantinova_nonequilibrium_2018}. A widely used approach to describe these dynamics is the effective temperature model, where the evolution of each (sub-)degree of freedom is characterized by its own effective temperature \cite{parker_modified_1975,allen_theory_1987,perfetti_ultrafast_2007}.
\par
In the case of superconductors, while only a few studies exist for conventional superconductors \cite{demsar_non-equilibrium_2020}, most studies of the superconducting (SC) to normal metallic state phase transition have focused on the cuprates, which have emerged as a test-bed for this particular class of photoinduced phase transitions \cite{giannetti_ultrafast_2016}. Early pump-probe optical reflectivity studies revealed that the critical pump fluence $F_c$ to destroy the SC state corresponds to a deposited energy by the pump pulse that is significantly below the thermal energy required to reach the SC critical temperature $T_c$. This finding points to the non-thermal nature of the phase transition where different degrees of freedom, here electron quasiparticles, phonons, or magnons, exhibit distinct transient dynamics in the photoinduced normal metallic state  \cite{kusar_controlled_2008,giannetti_discontinuity_2009,coslovich_evidence_2011,madan_dynamics_2017}. A similar conclusion was reached by time-resolved THz studies by tracking the photoinduced suppression of the superfluid density \cite{carnahan_nonequilibrium_2004,beyer_photoinduced_2011,beck_energy_2017}.
\par Deeper insights were provided by time-resolved Angle-Resolved-Photo-Emission Spectroscopy (TR-ARPES) which enabled direct tracking of the SC state collapse through the dynamics of the quasiparticle (QP) spectrum at the SC gap energy scale \cite{cortes_momentum-resolved_2011,smallwood_tracking_2012,smallwood_nonequilibrium_2016,miller_photoinduced_2015, piovera_quasiparticle_2015,ishida_quasi-particles_2016,zhang_photoinduced_2017,parham_ultrafast_2017,boschini_collapse_2018,zonno_time-resolved_2021,armanno_direct_2025}. In the cuprate Bi$_2$Sr$_2$CaCu$_2$O$_{8+\delta}$ (Bi2212) a sub-picosecond (ps) SC gap filling was observed in the near-nodal region of the Brillouin zone. This filling was found to be consistent with the equilibrium temperature-dependent ARPES data \cite{kondo_point_2015}, and was interpreted qualitatively as a signature of transient thermal heating of the electronic sub-system \cite{ishida_quasi-particles_2016}. The electron-only thermal picture was put on firmer grounds by Parham et al. who could describe the entire ps dynamics of TR-ARPES spectra of Bi2212 across the phase transition using a single effective electron temperature determined experimentally at the SC gap nodes \cite{parham_ultrafast_2017}. This electron-only thermal view was challenged by Zonno et al. who found a dichotomy between the nodal electron temperature and the pair-breaking scattering rate extracted from phenomenological fits of the SC QP spectral function \cite{zonno_time-resolved_2021}. The latter, assigned to SC coherence, exhibits faster dynamics compared to the electron temperature, which was interpreted as evidence for pump-induced phase fluctuations destroying SC coherence in a non-thermal fashion \cite{boschini_collapse_2018}. While the disagreement between these studies may emanate from differences in the modeling of the ARPES data, their emphasis on the near-nodal region highlights the need for further exploration of the dynamics towards the anti-nodal regions of the Brillouin zone. TR-ARPES studies spanning the entire Brillouin zone could not fully resolve the nature of the SC to normal phase transition due to their limited energy resolution \cite{dakovski_quasiparticle_2015,cilento_dynamics_2018}. 
\par
Here, we present an alternative point of view by focusing on the out-of-equilibrium dynamics of the anti-nodal region across the SC to normal phase transition through Time-Resolved Raman scattering spectroscopy (TR-Raman) on the cuprate Bi2212. 
We track the photoinduced SC condensate destruction and recovery with sub-ps time resolution using the Raman Cooper pair-breaking peak intensity as a marker of the SC coherence. Comparing its dynamics with the QP temperature obtained via the anti-Stokes part of the Raman spectrum, we uncover a significant dichotomy between the SC condensate and the QP temperature dynamics near the anti-nodes. We show that this difference cannot be framed in terms of a single effective electron temperature. Our data indicate that both SC destruction and its recovery over several ps are strongly non-thermal even within the electron sub-system. The present work demonstrates the ability of TR-Raman to selectively probe out-of-equilibrium pathways of different electron sub-degrees of freedom during a photoinduced phase transition.

\begin{figure*}[t]
    \centering
    \includegraphics[width=18cm]{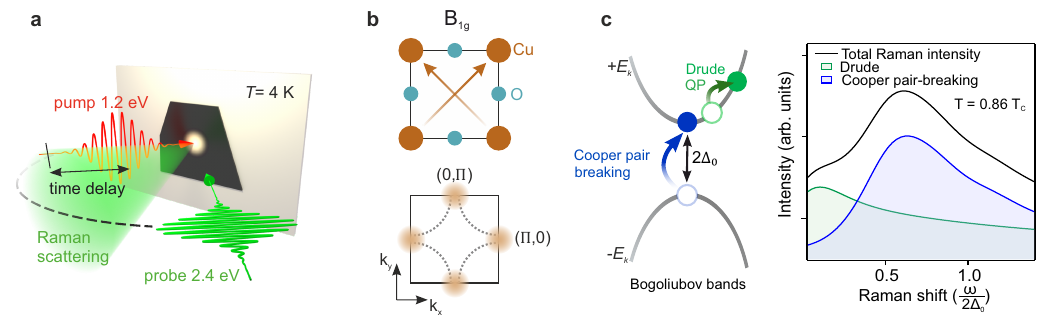}
    \caption{(a) Sketch of the two-color pump-probe set-up for TR-Raman spectroscopy. (b) The cross-polarization configuration corresponding to the Raman $B_{1g}$ channel is shown (top). This configuration selectively probes the anti-nodal regions of the Fermi surface of cuprates, where the d-wave SC gap is maximum (brown regions, bottom). (c) Illustration of the Cooper pair-breaking and quasi-particle (QP) Drude processes that contribute to the Raman spectrum in the SC state. From the point of view of the SC Bogoliubov bands ($+E_k$ and $-E_k$), they correspond to inter- and intraband processes, respectively. A theoretical Raman intensity spectrum in the $B_{1g}$ channel for a d-wave SC gap is shown for a temperature of 0.86$T_c$. $\Delta_0$ is the anti-nodal spectral gap at T=0~K. The Drude and pair-breaking contributions are shown in green and blue, respectively. The total Raman intensity is displayed in black (see the appendix for details of the calculations).}
    \label{fig1}
\end{figure*}

\section{Methods}

We performed TR-Raman measurements on a freshly cleaved optimally doped Bi-2212 single crystal with $T_c$ = 90~K. The sample was excited with a Yb-based amplified laser system emitting 150~fs linearly polarized laser pulses with a repetition rate of 250~kHz and a central wavelength of 1030~nm (1.2~eV). The estimated absorbed pump laser fluences used in this study range from 2 to 50~$\mu J.cm^{-2}$. To spectrally separate the pump and probe signals, the probe is set at 515~nm
(2.4~eV), produced via second-harmonic generation in a BBO crystal. Using the available optical constant data of optimally doped Bi2212 \cite{molegraaf_signposts_2005}, the pump and probe penetration depths were estimated to be 150~nm and 90~nm, respectively. Pump and probe beams were focused on the crystal in a non-collinear configuration (see Fig. \ref{fig1}(a)). The Raman probe signal was acquired using a single-grating spectrometer equipped with a nitrogen-cooled CCD camera. The elastic stray light was filtered out using wavelength-tunable long-pass (for Stokes spectra) and short-pass (for anti-Stokes spectra) edge filters.

A spectral pulse shaper similar to that presented in Ref.\cite{versteeg_tunable_2018}, was implemented in the probe beam path to obtain the best compromise between the time and energy resolutions. The overall time resolution was determined using the rise time of the photo-induced reflectivity variations, $\Delta R$, in the normal metallic state ($T > T_c$). In the present study, we used two settings of the probe spectral pulse shaper corresponding to the overall time resolutions of 0.36~ps and 0.43~ps. The Bi2212 crystal was mounted on the cold finger of a closed-cycle cryostat with a base temperature of 4~K. 
\par 
 All the Raman measurements reported here were performed in the $B_{1g}$ symmetry channel obtained using crossed polarizations between incident and scattered probe photons oriented at 45 degrees with respect to Cu-Cu bonds. This geometry selects the anti-nodal regions of the Fermi surface where the d-wave gap $\Delta_k$ is maximum \cite{devereaux_inelastic_2007} (see Fig. \ref{fig1}(b)). In the SC state, the Raman spectrum is dominated by the Cooper pair-breaking peak which corresponds to an interband excitation between the SC Bogoliubov bands at $\pm E_k=\sqrt{\epsilon_k+\Delta_k}$ where $\epsilon_k$ is the normal state electronic dispersion. In the $B_{1g}$ Raman spectrum of cuprates, this results in a pair-breaking peak located at twice the maximum of the spectral gap $2\Delta_0$ \cite{devereaux_inelastic_2007} (Fig. \ref{fig1}(c)). At finite temperature, thermally activated QPs will contribute as an additional Drude-like component in the spectrum, which corresponds to ungapped intra-Bogoliubov bands processes. Pair-breaking and QP Drude processes are sketched in Fig. \ref{fig1}(c). A theoretical equilibrium $B_{1g}$ intensity spectrum for a d-wave SC gap, highlighting pair-breaking and thermal QP Drude contributions, is shown in Fig. \ref{fig1}(c) (see appendix for details of the calculations). These two features of the Raman spectrum will serve as markers of the SC condensate and the QP Drude in our experimental spectra, as described in the following.

\section{Results}

\begin{figure}[htbp]
    \includegraphics[width=\columnwidth]{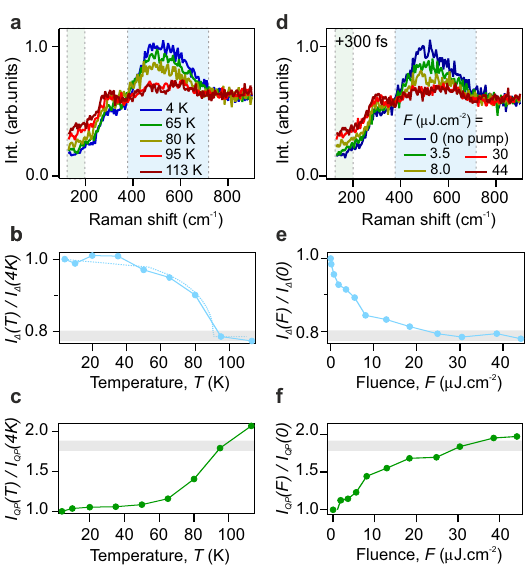}
    \caption{(a) Temperature-dependent Raman spectra in the $B_{1g}$ channel. The spectral regions used to define $I_{\Delta}$ and $I_{QP}$ are indicated in blue and green shaded regions, respectively. In between these regions, a phonon peak is observed at $\sim$ 260 $cm^{-1}$ (b) Evolution of spectral weights of the pair-breaking peak, $I_{\Delta}(T)$, and (c) low-energy Drude QPs, $I_{QP}(T)$, as a function of temperature. The integrals are divided by their corresponding values at 4~K. The horizontal grey area marks their respective value at $T=T_c$, accounting for error bars which can be used to define the critical fluence $F_T$. $I_{\Delta}$(T) can be described by the sum of a temperature-independent constant and a mean-field BCS temperature-dependent term (dashed line in panel b). (d) Pump-probe Raman spectra as a function of pump fluence for a fixed pump-probe delay of 300~fs and cold finger temperature of 4~K. (e, f) Evolution of $I_{\Delta}(F)$ and $I_{QP}(F)$ normalized to their pump-off values which are noted as $I_{\Delta}(0)$ and $I_{QP}(0)$, respectively.}
    \label{fig2}
\end{figure}

\subsection{Superconducting state destruction and recovery dynamics}
The equilibrium raw Raman intensity as a function of the Raman shift for cold finger temperatures ranging from ~4 K to 113~K is displayed in Fig. \ref{fig2}(a). At low temperature, the spectrum is dominated by the pair-breaking peak located at $\sim$ 500~ cm$^{-1}$ (blue shaded area). Upon increasing temperature, we observe a gradual spectral weight transfer from the pair-breaking peak towards lower energies, below $\sim$ 300~cm$^{-1}$, which is assigned to the emerging Drude component from thermally activated QPs. The nearly flat normal state spectrum has been assigned to marginal Fermi-liquid-like behavior of QPs with a linear frequency-dependent scattering rate \cite{devereaux_inelastic_2007}. Next, we investigate the effect of the pump fluence $F$ on the $T$ = 4~K Raman spectrum for a fixed pump-probe delay of +300~fs. Figure \ref{fig2}(d) demonstrates a reduction of the pair-breaking peak upon increasing pump fluence, resulting in its complete destruction for a fluence of $\sim$ 30~$\mu$J/cm$^{2}$. A more quantitative analysis of temperature and pump-induced effects on the Raman spectrum can be obtained by monitoring the quantity $I_\Delta$ defined as the integral below the pair-breaking peak, between 370 and 725~cm$^{-1}$, a spectral region which contains most of its spectral weight. Theoretically, the integral of the pair-breaking component of the Raman intensity is proportional to the coherent Cooper pair density weighted by the Raman vertex $\gamma_k$ 
\begin{equation}
I_{\Delta}\sim \sum_k \gamma_k^2 (u_kv_k)^2
\end{equation}
where $u_k$ and $v_k$ are the Bardeen-Cooper-Schrieffer (BCS) coherence factors and $(u_kv_k)^2=\frac{\Delta_k^2}{E_k^2}$. For a d-wave gap, $\Delta_k=\Delta_0 (cos(k_x)-cos(k_y))$, thus $I_{\Delta}\propto \Delta_0$ making $I_{\Delta}$ a marker of the coherent SC order parameter. Here we make a clear distinction between SC order parameter and SC pairing gap, which, while equivalent quantities within the BCS mean-field theory, may exhibit different behaviors in the case of cuprates. This is evidenced by the lack of significant softening of the pair-breaking peak energy in the temperature-dependent Raman spectra \cite{blanc_loss_2010}, and the gap filling rather than softening observed in ARPES data \cite{kondo_point_2015}  (see Appendix A for a more detailed discussion of the comparison between Raman and ARPES measurements). In practice, the integral $I_{\Delta}$ will also contain the high-energy tail of the Drude component, which is only weakly temperature dependent above $T_c$ and can be taken into account by a constant. An equivalent quantity $I_{QP}$ can be defined for the QP Drude component by integrating the low-energy part of the spectrum below the pair-breaking peak, between 130 and 200~cm~$^{-1}$. Figure \ref{fig2} shows the two integrals, renormalized by their values at $T$ = 4 K and $F$=0 $\mu$J/cm$^{2}$, respectively, as a function of temperature (panel b and c) and fluence (panel e and f). Above $T_c$, $I_{\Delta}(T)$ is essentially temperature-independent \cite{blanc_loss_2010} while below $T_c$, it follows a mean-field-like temperature dependence, confirming $I_{\Delta}(T)$ as a marker of the SC order parameter. By contrast, $I_{\Delta}(F)$ exhibits a pronounced fluence-dependent behavior; its strong sensitivity is evidenced by a significant intensity loss even at low pump fluence, highlighting the large excess energy delivered by the pump photons relative to thermal excitation. A similar contrast between thermal and photoinduced effects is observed for the QP Drude component $I_{QP}$  (Figs. \ref{fig2} (e, f)). The threshold pump fluence $F_{T}$ to fully destroy superconductivity, defined as  
$I_{\Delta}(F_T,T=4~K)$=$I_{\Delta}(F=0,T=T_C)$, is 25$\pm 5 ~\mu J.cm^{-2}$, in overall agreement with previous estimates on Bi2212 \cite{carnahan_nonequilibrium_2004,coslovich_evidence_2011,smallwood_tracking_2012,zhang_photoinduced_2017,parham_ultrafast_2017,boschini_collapse_2018}. 

\begin{figure*}[]
    \includegraphics[width=18.0cm]{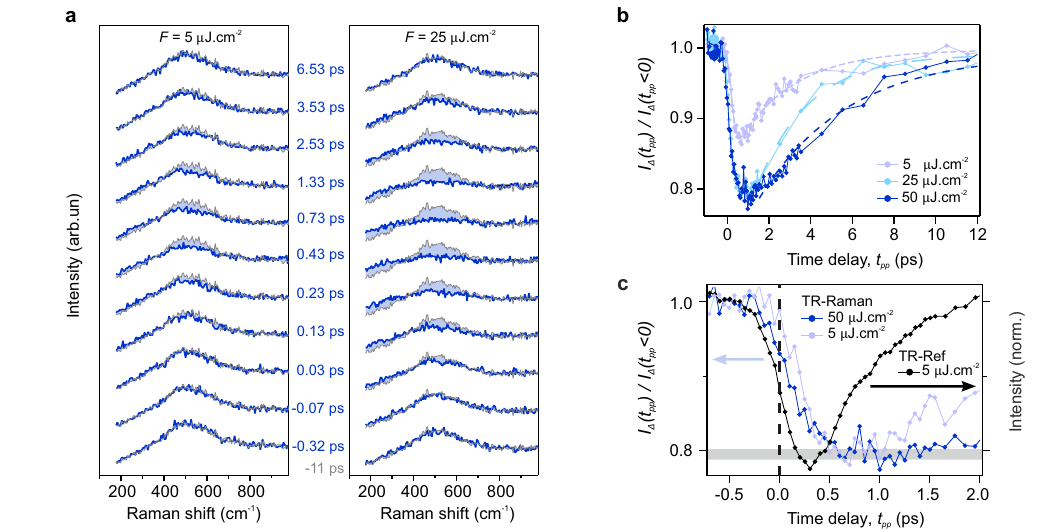}
    \caption{ (a) TR-Raman spectra at selected pump-probe delays for two different pump fluences, 5~$\mu J.cm^{-2}$ and 25~$\mu J.cm^{-2}$, (blue curves). The negative delay spectrum at -11 ps (grey curve) is overlaid on the data to highlight the pump-induced changes. (b) Dynamics of the pair-breaking peak intensity, $I_{\Delta}(t_{pp})$, for three different pump fluences as a function of the pump-probe time delay, $t_{pp}$. The data are normalized to the negative delay spectra noted as $I_{\Delta}(t_{pp}<0)$. Dashed lines show the exponential fits for the decay dynamics. (c) Magnified view of the same dynamics near the zero time delay (left axis). The shaded grey area marks the plateau region. The time-resolved reflectivity (TR-Ref) trace in the normal state ($T$ = 113~K) using the same pump-probe conditions is also shown for comparison (right axis).}
    \label{fig3}
\end{figure*}

Figure \ref{fig3} presents TR-Raman data for two different pump fluences corresponding to $F<F_T$ ($5~\mu J.cm^{-2}$) and $F \sim F_T$ (25~ $\mu J.cm^{-2}$) at selected pump-probe time delays $t_{pp}$. For these scans, a time resolution of 0.36~ps was selected. To highlight pump-induced changes in the spectra, the negative delay spectrum at -11~ps is superimposed in grey for each time delay. The TR-Raman spectra demonstrate a sub-ps weakening/destruction of the pair-breaking peak and its recovery within a few picoseconds. The pair-breaking peak temporal dynamics is accompanied by the emergence and the subsequent decay of a low energy ($<$ 300~cm$^{-1}$) QP Drude component, qualitatively resembling the thermal evolution observed in equilibrium Raman spectra (Fig. \ref{fig2} (a)). These observations are consistent with a previous TR-Raman study, which was performed with lower time resolution on an overdoped Bi2212 sample \cite{saichu_two-component_2009}. The dynamics of the SC order parameter can be monitored as a function of pump-probe delay $t_{pp}$ using $I_{\Delta}(t_{pp})$/$I_{\Delta}(t_{pp}<0)$ where $I_{\Delta}(t_{pp}<0)$ is its negative delay value. The integrated intensity of the pair-breaking peak is shown in Fig. \ref{fig3}(b) for various pump fluences. At low fluence $F<F_T$, the depletion and recovery of SC state can be well fitted using a single exponential, $Ae^{-t_{pp}/\tau}$, convoluted with a Gaussian function representing the finite-time resolution. At higher fluences, the exponential decay response is delayed and is preceded by a plateau-like response between 0.5 and 1.5~ps. This behavior can be interpreted as the complete destruction of the SC state that persists for approximately 1~ps (see Fig. \ref{fig3}(c)). Fitting the data at high fluence with an exponential decay function starting where the plateau terminates ($t_{pp}$=1.5~ps), we obtain a SC recovery time $\tau$ of about 2.8~ps for both $F< F_T$ and $F \sim F_T$, and 4.8~ps for $F> F_T$.
%These SC recovery times are close to the one reported by previous pump-probe measurements in Bi2212 %\cite{coslovich_evidence_2011,smallwood_time-_2014,zhang_photoinduced_2017,madan_dynamics_2017}. 

\par
Our data also reveal interesting insights into the early time dynamics of the SC destruction process. In Fig. \ref{fig3} (c), we compare the near-zero-delay dynamics of the pair-breaking Raman signal ($I_{\Delta}(t_{pp})$) with the transient reflectivity changes in the normal state ($\Delta R$) measured in the same pump-probe geometry. The rationale for using $\Delta R$ data in the normal state as a reference lies in its sensitivity to the sub-0.1~ps heating of the normal state electrons, and thus will be limited by our lower time resolution \cite{perfetti_ultrafast_2007}. The comparison reveals a small but distinct delay time of $\sim$0.2~ps for the onset of pump-induced destruction of the SC state with respect to time-resolved reflectivity data. The delay is present at all fluences but is reduced with increasing fluence. Overall, this delay indicates that the pump-induced destruction of superconductivity does not primarily occur via direct pump-induced Cooper pair-breaking processes but is instead mediated by bosonic excitations, i.e., magnons or phonons, generated during the initial cooling of photoexcited hot electrons \cite{demsar_non-equilibrium_2020}. In this picture, the early-stage dynamics is governed by the interplay between the superconducting condensate and the pair-breaking bosons \cite{rothwarf_measurement_1967,kabanov_kinetics_2005,kusar_controlled_2008,demsar_non-equilibrium_2020}. Interestingly, theoretical models of this coupled dynamics using Rothwarf-Taylor equations predict a fluence-dependent increase in SC destruction time, consistent with our observations \cite{demsar_pair-breaking_2003,kabanov_kinetics_2005,demsar_non-equilibrium_2020}. Thus, our TR-Raman measurements of antinodal dynamics support a picture of indirect, boson-mediated SC condensate suppression.

\subsection{Raman thermometry of the SC out-of-equilibrium dynamics}

\begin{figure*}[]
   \includegraphics[width=18.0cm]{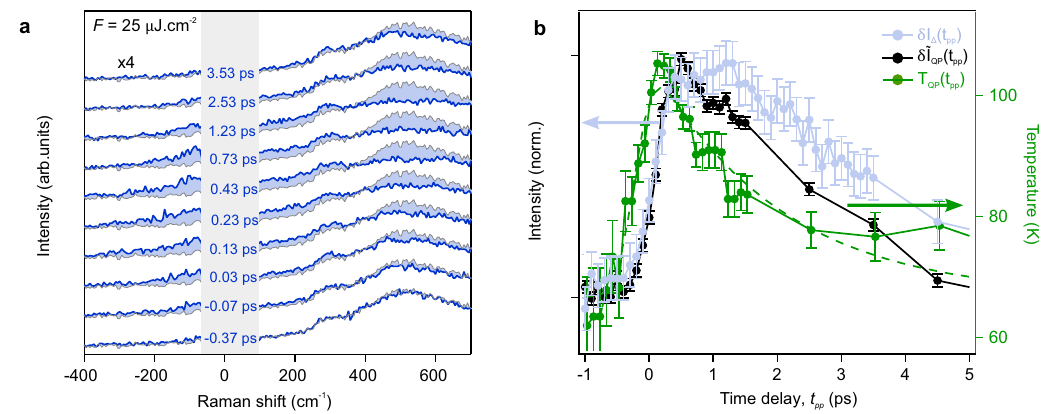}
    \caption{(a) Anti-Stokes (AS) and Stokes (S) TR-Raman spectra for $F\sim F_T$ at selected pump-probe delays. The AS spectra have been multiplied by 4 for clarity. The edge filter's cut-offs at the center of the spectra are covered by the grey shaded zone. (b) Comparison between the dynamics of $T_{QP}(t_{pp})$ extracted from the AS and S spectra and the change in $I_{\Delta}(t_{pp})$ and $\tilde I_{QP}(t_{pp})$ with respect to negative delays, $\delta{I_{\Delta}}(t_{pp})$ and $\delta \tilde{I}_{QP}(t_{pp})$, respectively. See text for the definition of $\tilde I_{QP}(t_{pp})$. The data are normalized to their maximum values. The intensity of $\delta{I_{\Delta}}(t_{pp})$ and $\delta \tilde{I}_{QP}(t_{pp})$ is plotted on the left axis while the temperature is shown on the right axis. In this figure error bars have been included for the integrated intensities. They are based on the statistical analysis of the intensity changes of the Raman spectra at high energy ($>$1500~cm$^{-1}$) attributed to small changes in the laser position during pump-probe scans.}
    \label{fig4}
\end{figure*}

We now address the question of the nature of the photoinduced phase transition between the SC and the normal state by combining the previous results with ultrafast Raman thermometry. In Fig. \ref{fig4}(a), we report a series of TR-Raman spectra of both anti-Stokes (AS, $\Omega <$ 0) and Stokes (S, $\Omega >$ 0) processes for a pump fluence close to $F_T$. To access lower Raman shifts on both the Stokes and anti-Stokes sides, the measurements were performed with higher spectral resolution, i.e., $\sim$ 60~cm$^{-1}$, resulting in a reduced overall time resolution of $\sim$ 0.43~ps. In equilibrium and at low temperature, $k_BT < \lvert\Omega\rvert$, we expect a vanishingly small AS signal due to the absence of significant thermal excitations, as observed for the negative delay AS spectra. However, a clear enhancement of AS spectra is observed above $\sim$-300~cm$^{-1}$ upon photoexcitation. As stated above, this spectral region is dominated by the QP Drude contribution (c.f. Fig. \ref{fig2} a). Since these QPs thermalize among themselves on a much faster time scale ($<$0.1~ps) than our time resolution \cite{perfetti_ultrafast_2007,ishida_quasi-particles_2016}, the dynamics of their contribution to Raman spectra can be described by an effective temperature, $T_{QP}$, which can be obtained via the ratio of AS and S Raman intensity. Indeed, the intensities of the AS and S contributions are linked to the Raman response function $\chi''$ via the Bose-Einstein distribution function $n_B$: $I_{S}(\Omega)\propto (1+n_B)(\Omega,T)\times\chi''(\Omega)$ and $I_{S}(-\Omega)\propto  n_B\times(\Omega,T)\chi''(\Omega)$. This results in the detailed-balance relation: $\frac{I_S}{I_{AS}}=e^{\beta \Omega}$ where $\beta=\frac{1}{k_BT_{QP}}$. For a pulsed laser this standard expression needs to be modified to account for the finite linewidth of the probe pulse leading to \cite{shvaika_interpreting_2018,matveev_stroboscopic_2019}:
\begin{equation}
    \frac{I_S(\tilde{\Omega}-\frac{\beta}{\sigma^2_b})}{I_{AS}(-\tilde{\Omega}-\frac{\beta}{\sigma^2_b})} = e^{\beta \tilde{\Omega}}
    \label {AS-S}
\end{equation}
where $\tilde{\Omega} = \Omega - \frac{\beta}{\sigma^2_b}$, $\Omega > 0$, and $\sigma_b$ is the Gaussian linewidth of the probe pulse amplitude in the time domain. This expression differs from the standard one through the temperature-dependent energy shift $\frac{\beta}{\sigma^2_b}$, which cannot be ignored for sub-ps probe pulses. Equation \ref{AS-S} was solved numerically for $T_{QP}(\Omega)$ which was then averaged over the energy range of $\Omega$ = 100 - 150~cm$^{-1}$. The lower bound is the lowest energy measured on the Stokes side, and the upper bound was chosen to avoid any overlap with the phonon at 280~cm$^{-1}$ and to have sufficient AS Raman signal at all delays to obtain reliable results (see Appendix B for details). The resulting $T_{QP}$ dynamics is shown as a function of pump-probe time delay, $t_{pp}$, in Fig. \ref{fig4}(b) (green curve). 
\par
Close to zero delay, $T_{QP}$ exhibits a sharp resolution-limited increase from $\sim$ 60~K at negative delays to $\sim$ 105~K close to $t_{pp} \sim$ 0~ps, qualitatively consistent with an ultrafast transition to the normal state. The relatively high temperature at negative delays compared to the cold finger temperature is ascribed to a combination of pump-induced average heating and possible self-pumping effects from the probe. The overall dynamics of $T_{QP}(t_{pp})$ can be well described by a single exponential recovery with a 2~ps characteristic time constant convoluted with a Gaussian time-resolution function. We did not find any evidence for an anomalously large anti-Stokes signal above -300~cm$^{-1}$ that would violate the principle of detailed balance, as recently reported by Glier et al. \cite{glier_non-equilibrium_2025} and interpreted as pump-induced generation of metastable SC Higgs mode excitations. This would have resulted in a significantly larger $T_{QP}$ in our energy window and, presumably, a much slower recovery time than the $\sim$2~ps observed here. Since Glier et al. report TR-Raman data only for a single pump-probe delay of 3~ps, using significantly broader pump and probe pulses ($\geq$ 1~ps) and higher pump fluences, a direct comparison with our data is not straightforward.

\par The ultrafast dynamics of $T_{QP}(t_{pp})$, can be compared with that of the SC condensate and the QP Drude discussed in the previous section. To this end, the evolutions of the SC order parameter defined via $\delta I_{\Delta}(t_{pp})=I_{\Delta}(t_{pp})-I_{\Delta}(t_{pp}<0)$ and the QP Drude weight via $\delta \tilde I_{QP}(t_{pp})= \tilde I_{QP}(t_{pp})- \tilde I_{QP}(t_{pp}<0)$ are normalized to their maximum value and plotted in Fig. \ref{fig4} (b). The $\delta I_{\Delta}(t_{pp})$ curve is multiplied by -1 for direct comparison. Here, we note that the QP Drude weight $\tilde I_{QP(t_{pp})}$ is defined as the integral of the difference between the S and AS spectra $I_{QP, S}(\Omega)-I_{QP, AS}(-\Omega$). As this quantity does not depend on the Bose population factor $n_B(T_{QP})$, it better reflects the QP Drude weight, and hence the QP population, encoded in the Raman response function $\chi''$ than $I_{QP}(F)$ shown in Figs. \ref{fig2}(e, f).

\begin{figure}[htbp]
    \includegraphics[width=\columnwidth]{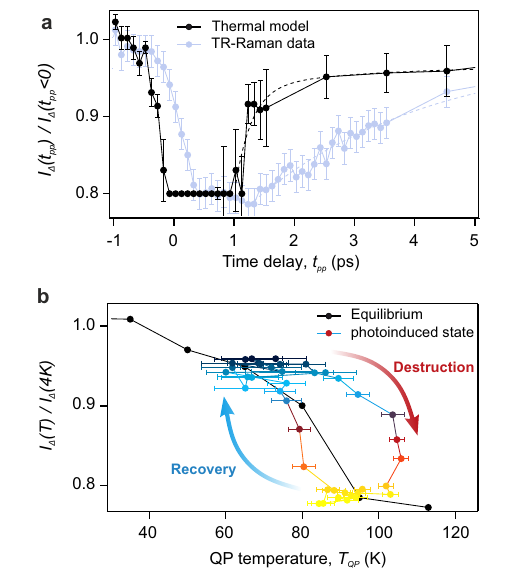}
    \caption{(a) Comparison between the dynamics of $I_{\Delta}(t_{pp})$ (pale blue) measured with TR-Raman spectroscopy and the one expected from the thermal ($T^*$) model $I^{T_{QP}}_{\Delta}(t_{pp})$ (black curve, see text for the description). (b) Similar comparison but as a function of $T_{QP}$. The temperature dependence of the equilibrium state is shown in black while the transient one, $I^{T_{QP}}_{\Delta}(t_{pp})$, is shown on a red-to-blue color scale. The SC destruction and recovery transient pathways are distinct and do not follow the one expected for the effective electron thermal model.}
    \label{fig5}
\end{figure}

In short pump-probe delays, there is a clear shift between the dynamics of $T_{QP}(t_{pp})$ and both $\delta\tilde I_{QP}(t_{pp})$ and $\delta I_{\Delta}(t_{pp})$. While $T_{QP}(t_{pp})$ reaches values above equilibrium $T_c$ quasi-instantaneously within our time-resolution, the SC condensate remains initially largely un-depleted. The shift of about 0.2~ps is similar to the one reported in Fig.\ref{fig3}(c). The SC state is fully destroyed only after 0.3~ps where both $\delta\tilde I_{QP}(t_{pp})$ and -$\delta I_{\Delta}(t_{pp})$ have reached their maxima. Both quantities follow each other over during first 0.5~ps following laser excitation, indicating that the increase of the QP Drude weight stems from a photoinduced population of QPs which were initially in the SC condensate. Between 0.3~ps and 1.5~ps $\delta I_{\Delta}(t_{pp})$ is essentially flat, suggesting a fully destroyed SC state as already stressed above. Conversely, the QP Drude weight $\delta\tilde{I}_{QP}(t_{pp})$ decreases during this time, an effect which cannot be attributed to QP recombination into Cooper pairs since $\delta I_{\Delta}(t_{pp})$ remains unchanged and flat. The faster decay of $\delta\tilde{I}_{QP}(t_{pp})$ likely reflects the relaxation of QPs through a transient pseudogap state, as the latter is known to induce a low energy depletion of the $B_{1g}$ Raman spectrum at thermal equilibrium upon lowering temperature from the normal state \cite{venturini_observation_2002,sacuto_new_2013}. Focusing on $\delta I_{\Delta}(t_{pp})$ and $T_{QP}(t_{pp})$, it is clear from Fig. \ref{fig4} (b) that they exhibit distinct dynamics both in the initial destruction and recovery stages. Since the SC order parameter and the QP temperature are directly tied via the BCS gap equation in thermal equilibrium, our data suggest that the destruction and recovery dynamics of the photoexcited SC condensate and the QPs in the anti-nodal regions of the Brillouin zone cannot be described by a single effective electron temperature.

\par
Further evidence for the non-thermal nature of the out-of-equilibrium dynamics in the electron subsystem comes from comparing the SC order parameter dynamics obtained from TR-Raman measurements, $I_{\Delta}(t_{pp})$, with that expected under a thermal scenario. In the thermal picture that is commonly referred to as the $T^*$ model in the context of non-equilibrium superconductivity, one assumes that for each pump-probe delay the electronic state can be fully described by a single effective electronic temperature, $T_{QP}$, \cite{parker_modified_1975,nicol_comparison_2003}.
This allows us to construct a thermal reference for the pair-breaking peak intensity, $I^{T_{QP}}_{\Delta}(t_{pp})$, by replacing the cold finger temperature $T$ in the equilibrium data (see Fig. \ref{fig2} (a, b)) with the transient QP temperature, $T_{QP}(t_{pp})$, at every delay as extracted for each delay from the Stokes-anti-Stokes analysis (green curve, Fig. \ref{fig4} (b)). Figure \ref{fig5} (a) show that the measured quantity $I_{\Delta}(t_{pp})$ in blue curve deviates significantly from $I^{T_{QP}}_{\Delta}(t_{pp})$ in black. The flat part of the curve $I^{T_{QP}}_{\Delta}(t_{pp})$ corresponds to the absence of superconductivity in the corresponding time interval. This non-thermal behavior is further highlighted in Fig. \ref{fig5} (b), where plotting $I_{\Delta}(t_{pp})$ versus $T_{QP}$ reveals a clear deviation of the photoinduced trajectory from the purely thermal path during both the destruction and recovery phases.

\section{Discussion}
The emerging picture from our TR-Raman data in Figs. \ref{fig4} and \ref{fig5} is that of a significant dichotomy between the QP temperature $T_{QP}$ and the SC order parameter at the anti-nodes when the system is driven across the SC to normal state phase transition. Based on our data and previous studies, the SC state dynamics in the anti-nodal region can be summarized into four different phases. (1) During the first 0.1~ps, the pump pulse generates a population of hot electrons which quickly thermalize among themselves, resulting in a resolution-limited rise of $T_{QP}$. These hot electrons likely originate from thermally activated QPs already present in equilibrium, and therefore do not initially cause a significant SC depletion. (2) During the next few hundred femtoseconds, the hot QPs cool down by emitting bosons (phonons or magnons) with energies close to twice the anti-nodal SC gap energy scale $2\Delta_0$, breaking Cooper pairs and creating low-energy QP near the gap edge. In this regime, there is a concomitant depletion of the SC condensate as observed in $I_{\Delta}(t_{pp})$ and the creation of a population of low-energy QPs, which will contribute to an increase of the QP Drude weight $\tilde{I}_{QP}(t_{pp})$. Within this picture, the observed $\sim$ 0.2~ps delay for the SC destruction can be ascribed to the coupled dynamics of pair-breaking bosons and low-energy QPs described by Rothwarf-Taylor equations as discussed above. (3) Between 0.5 and 1.5~ps, the SC condensate is fully depleted, but the QPs continue to cool down without recombining into Cooper pairs. This results in a non-thermal state where a fully depleted condensate co-exists with a population of "cold" QPs, which have not yet recombined into Cooper pairs. This state bears some resemblance to the phenomenology of the $\mu^*$ model, a non-thermal framework which describes a cold population imbalance of QPs which reduces the SC order parameter and results from the mismatch between a fast electron thermalization time and a slower Cooper-pair recombination time \cite{owen_superconducting_1972,nicol_comparison_2003}. (4) The subsequent recovery dynamics of the SC state, lasting for a few ps, is governed by the recombination of cold QPs into Cooper pairs, a process that is mainly limited by the decay rate of pair-breaking bosons \cite{kabanov_kinetics_2005,demsar_non-equilibrium_2020}. 
\par
The reported transient discrepancy between the SC coherence embodied in $I_{\Delta}(t_{pp})$ and the QP temperature is, at first glance, consistent with the TR-ARPES data of Zonno et al. \cite{zonno_time-resolved_2021} mentioned earlier. However, in their case, the SC coherence probed by the pair-breaking scattering rate in the near-nodal region recovers faster than the nodal QP temperature, which is different from our findings. Taken at face value, the discrepancy could be attributed to a significant anisotropy in QP thermalization and/or coherent SC order parameter recovery, aspects on which no consensus has emerged yet \cite{cortes_momentum-resolved_2011,smallwood_tracking_2012,smallwood_time-_2014,gedik_diffusion_2003,parham_ultrafast_2017}. Indeed the nodal electron temperature relaxation time extracted from TR-ARPES ($\sim$ 3.7~ps \cite{parham_ultrafast_2017}) appears notably slower than the one extracted from our measurements near the anti-nodal region $\sim$2 ~ps. It is not clear whether the dynamics in the nodal and anti-nodal regions are indeed different, or whether our results point out a fundamental difference between the SC coherence probed by Raman and ARPES techniques. This point deserves further scrutiny.
 
\section{Conclusion}
In conclusion, we have used ultrafast TR-Raman scattering to monitor the anti-nodal SC dynamics of Bi2212 across the photoinduced SC to normal state phase transition. Analyzing the ultrafast evolution of the SC condensate and the QP temperature with the help of the pair-breaking peak and Raman thermometry, respectively, we found a significant dichotomy between the QP thermalization and Cooper-pair recombination dynamics, which cannot be accounted for by a thermal electronic picture. Our study highlights the potential of TR-Raman spectroscopy in the study of light-induced phase transitions in quantum materials, where different degrees of freedom can display distinct dynamics. The proposed approach, based on the spectral selectivity of Raman scattering and its ability to track excitation populations, is general, and we expect it to be applicable to other transient states whose non-thermal nature is often inferred only indirectly.

\section{Acknowledgments}
We thank I. Paul and L. Perfetti for insightful discussions. We acknowledge funding from the Agence National de la Recherche via the grants ANR "SUPER2DTMD" and ANR "TERADIRAC". This work has been supported by Region Île-de-France in the framework of DIM QuanTiP and DIM SIRTEQ. The work at BNL was supported by the US Department of Energy, office of Basic Energy Sciences, contract no. DOE-SC0012704.
\newpage
\section{Appendix A: BCS Raman response with finite lifetime effects}
The starting point in the general expression of the imaginary part of the Raman susceptibility in the BCS approximation:
\begin{multline}
\chi"(\omega)=\sum_k\gamma_k^2\int d\omega'(f(\omega')-f(\omega'+\omega))(A_{11}(\omega'+\omega)A_{11}(\omega') \\+A_{22}(\omega'+\omega)A_{22}(\omega')-2A_{12}(\omega'+\omega)A_{12}(\omega')
\label{chi}
\end{multline}
\cite{valenzuela_phenomenological_2007}
In equilibrium $\chi"$ is linked to the measured Stokes Raman intensity $I_R$ via the Bose occupation factor $n$:
\begin{equation}
I_R\propto (1+n(\omega,T)) \chi"
\label{bose}
\end{equation}

$A_{11}$ and $A_{22}$ are the normal spectral function and the $A_{12}$ are the anomalous spectral function. $\gamma_k$ is the Raman vertex whose k dependence will depend on light polarizations. For e.g. $\gamma_{B_{1g}}\propto cos(k_x)-cos(k_y)$.
Neglecting lifetime effects, the spectral functions are given by:

\begin{align}
A_{11}=\frac{\pi}{2E_k}(\omega+\epsilon_k)(\delta(\omega+E_k)-\delta(\omega-E_k)) \\
A_{22}=\frac{\pi}{2E_k}(\omega-\epsilon_k)(\delta(\omega+E_k)-\delta(\omega-E_k)) \\
A_{12}=\frac{\pi}{2E_k}\Delta_k(\delta(\omega+E_k)-\delta(\omega-E_k))
\end{align}
The pole of the normal spectral function corresponds to the Bogoliubov quasiparticles with energy $E_k=(\epsilon_k^2+\Delta_k^2)^{\frac{1}{2}}$. $\Delta_k$ is the BCS order parameter. Note that the order parameter appears both in the numerator of the anomalous spectral function and in the energy $E_k$.

The corresponding Raman $\chi"$ can be computed analytically:

\begin{equation}
\chi"(\omega>0)=\pi^2\sum_k th(\frac{E_k}{2k_BT})\frac{\Delta_k^2}{E_k^2}\delta(\omega-2E_k)
\end{equation}

For free quasiparticles, the Raman response is zero when $\Delta_k=0$: there is no response in the normal state, and also no response coming from thermally excited quasiparticles. The response is entirely given by the SC condensate: the Cooper-pair-breaking channel which creates a pair a Bogoliubov quasiparticles with energies $-E_k$ and $E_k$.  Note that the integral of the Cooper pair-breaking response over the frequency of the response is:
\begin{equation}
I_{\Delta}=\int \chi'' d\omega=\pi^2\sum_k\frac{\gamma_k^2\Delta_k^2}{E_k^2}
\end{equation}
The quantity $\frac{\Delta_k^2}{E_k^2}=4 \mid u_kv_k \mid^2$ is the condensed Cooper-pair density, which will be finite only below $T_c$ where a coherent SC condensate forms. This contrasts with the spectral energy gap, which does not soften significantly upon approaching $T_c$ in the $B_{1g}$ Raman spectra of cuprates. The sensitivity of Raman scattering to Cooper-pair density is associated with the fact that, being a two-particle response, it probes the anomalous BCS spectral functions which are finite only when long-range coherence settles. This contrasts with one-particle probe like ARPES measurements, which only probe the normal BCS spectral function, where the SC will manifest mainly through the spectral gap. While the spectral gap and the coherent pair formation density are governed by the same quantity $\Delta_k$ in mean-field BCS theory, this is not the case in cuprates, where the phase transition is likely not set by the spectral gap closing, but by the loss of phase coherence. 
\par
The dichotomy between the continuous collapse of the integrated quantity $I_{\Delta}(t_{pp})$ and the absence of significant softening of the pair-breaking peak energy, which measures the spectral gap, has already been documented in previous static Raman measurements in cuprates \cite{blanc_loss_2010}.  It is consistent with ARPES measurements where a gap filling rather than closing is observed \cite{kondo_point_2015}. 
\par

For an isotropic gap $\Delta_k=\Delta_0$ and at T=0K, the integral is proportional to the SC order parameter:
\begin{multline}
I_{\Delta}=\pi^2\Delta_0^2\int d\epsilon \sum_k \gamma_k^2\delta(\epsilon-\epsilon_k) \frac{1}{\epsilon^2+\Delta_0^2} \\
=\pi^2\Delta_0^2\int d\epsilon \phi(\epsilon)\frac{1}{\epsilon^2+\Delta_0^2}
\end{multline}
where $\phi(\epsilon)=\sum_k\gamma_k^2\delta(\epsilon-\epsilon_k)$ is the density of state weighted by the Raman vertex. Note that the fraction has a strong peak at $\epsilon=0$ so that assuming that the density of state does not vary too much close to the Fermi energy (near $\epsilon_k$=0) over an energy range on the order of $2\Delta_0$ we can replace $\phi(\epsilon)$ by $\phi(0)$. The remaining integral can then be computed analytically:
\begin{equation}
I_{\Delta}\propto \pi^3\phi(0)\Delta_0
\end{equation}
\par
For a d-wave gap with $\Delta_k=f_k\delta_0$ with the form factor $f_k=(cos(k_xa)-cos(k_ya)$ a similar expression can be obtained:
\begin{multline}
I_{\Delta}=\pi^2\Delta_0^2\int d\epsilon \sum_k \gamma_k^2\delta(\epsilon-\frac{\epsilon_k}{f_k}) \frac{\Delta_0^2}{\frac{\epsilon^2}{f_k^2}+\Delta_0^2}\\
=\pi^2\Delta_0^2\int d\epsilon \tilde\phi(\epsilon)\frac{1}{\epsilon^2+\Delta_0^2}
\end{multline}
where $\tilde\phi$ now involves the form factor $f_k$. Assuming that this quantity just like the DOS does not vary strongly close ti the Fermi energy we have:
\begin{equation}
I_{\Delta}\sim \pi^3\tilde\phi(0)\Delta_0
\end{equation}

\par
The previous response is due solely to the creation of a Bogoliubov quasiparticle (Cooper pair-breaking channel). There is no contribution from thermally excited quasiparticles in the SC state and thus no response in the normal state. The first step to describe the QP Drude response is to consider a non-zero scattering rate $\Gamma$ for the normal state electrons (due to for e.g. disorder). In that case the spectral functions need to be recomputed from the original Green's function $\hat{G}$ with finite scattering rate:
\begin{align}
A_{11}=\frac{\Gamma}{2E_k} \{\frac{E_k+\epsilon_k}{(\omega-E_k)^2+\Gamma^2}+\frac{E_k-\epsilon_k}{(\omega+E_k)^2+\Gamma^2}\} \\
A_{22}=\frac{\Gamma}{2E_k} \{\frac{E_k-\epsilon_k}{(\omega-E_k)^2+\Gamma^2}+\frac{E_k+\epsilon_k}{(\omega+E_k)^2+\Gamma^2}\}  \\
A_{12}=\frac{\Gamma\Delta_k}{2E_k} \{\frac{1}{(\omega-E_k)^2+\Gamma^2}-\frac{1}{(\omega+E_k)^2+\Gamma^2}\} 
\end{align}
When computing $\chi"$ using equation \ref{chi} we can group the terms into 4 types of integrands: 2 new "Drude" terms and 2 "SC" terms. We can  separate each contribution to $\chi''$:
\begin{equation}
\chi"=\chi^"_{SC}+\chi^"_{Drude}
\end{equation}
Using $E_k^2=\Delta_k^2+\epsilon_k^2$ we obtain the following expressions for the SC and Drude contributions:

\begin{multline}
\chi^"_{Drude}=\sum_k\frac{\gamma_k^2\Gamma^2\epsilon_k^2}{E_k^2} \int d\omega'((f(\omega')-f(\omega'+\omega)))\\
(\frac{1}{((\omega'-E_k)^2+\Gamma^2)((\omega+\omega'-E_k)^2+\Gamma^2)} 
\\
+\frac{1}{((\omega'+E_k)^2+\Gamma^2)((\omega+\omega'+E_k)^2+\Gamma^2)})
\end{multline}
This is very similar to the standard Raman QP Drude response and will give a overdamped Lorentzian centered around $\omega=2\Gamma$.
\begin{multline}
\chi^"_{SC}=\sum_k\frac{\gamma_k^2\Gamma^2\Delta_k^2}{E_k^2} \int d\omega'((f(\omega')-f(\omega'+\omega)))\\
(\frac{1}{((\omega'+E_k)^2+\Gamma^2)((\omega+\omega'-E_k)^2+\Gamma^2)} \\
\\+\frac{1}{((\omega'-E_k)^2+\Gamma^2)((\omega+\omega'+E_k)^2+\Gamma^2)})
\end{multline}
We numerically checked that the integrated response $I_{\Delta}$ of the pair-breaking SC contribution  $\chi''_{SC}$ still scales with $\Delta_0$ when lifetime effects are included: $I_{\Delta}\propto \Delta_0$.
The Raman intensity $I_R$ of each contribution at a given temperature can be obtained via equation \ref{bose}. The Raman intensity decomposed into QP Drude and pair-breaking contribution for T=0.86~$T_c$ and $\Gamma=\frac{\Delta_0(T=0K)}{10}$ for the $B_{1g}$ symmetry channel is shown in the figure \ref{fig1} of the main text. Since to our knowledge, there is no satisfactory framework to separate out the coherent SC order parameter and the pairing gap in describing the Raman response of cuprates, we took a single $\Delta_k$ as in standard BCS theory. The theoretical spectrum, therefore, includes a significant softening of the spectral energy gap, which can be read off from the pair-breaking peak energy, at a temperature close to $T_c$.
The QP Drude part coming from thermally excited QPs is confined to energies well below the SC gap, while the pair-breaking contribution is centered at higher energies around twice the BCS spectral gap. The relatively clear spectral separation between the 2 contributions allows us to separate them without relying on any fit of the actual data.

\section{Appendix B}
By measuring the Stokes and Antistokes Raman signals, we have access to the population of the excitations, which is linked to their temperature. Using the principle of detailed balance, we have the analytical equation:

\begin{equation}
    \frac{I_S(\Omega)}{I_{AS}(-\Omega)} = e^{\beta\Omega}.
\end{equation}
Where $\mathrm{I}_S$ and $\mathrm{I}_{AS}$ stand for the Stokes and Antistokes intensities respectively, $\Omega$ is the Raman shift and $\beta = 1/k_BT$ is the Boltzmann factor. However, this formula is valid for data acquired via a continuous laser with essentially zero energy linewidth. For a pulsed laser probe with a significant energy linewidth, the previous formula has to be corrected via an effective energy shift of the Raman shift $\Omega$ which depends on both the temperature and the laser linewidth. The demonstration of the formula can be found in \cite{matveev_stroboscopic_2019}. It reads:
\begin{equation}
    \frac{I_S(\Omega-\frac{2\beta}{\sigma_b^2})}{I_{AS}(-\Omega)} = e^{\beta(\Omega-\frac{\beta}{\sigma_b^2})}.
    \label{eq_corr}
\end{equation}
The energy shift $\frac{\beta}{\sigma_b^2}$  depends on both the temperature $T$ and the energy linewidth of the laser $\frac{1}{\sigma_b}$. With our laser linewidth we have $\frac{1}{\sigma_b}=$13~cm$^{-1}$ which gives an energy shift  $\frac{\beta}{\sigma_b^2}\sim$2.8~cm$^{-1}$ at 90~K. 
\par
\begin{figure}[h!]
    \includegraphics[width=8.5cm]{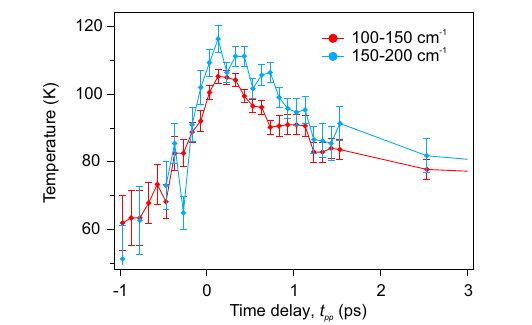}
    \caption{Comparison between the temperature extracted on two energy ranges. In red, between 100 and 150~$\mathrm{cm}^{-1}$, in blue between 150 and 200~$\mathrm{cm}^{-1}$. Temperature extracted from the higher energy range shows slightly higher temperatures close to zero delay.}
    \label{fig7}
\end{figure}

 
To extract the temperature, we define the function $y$ as:
\begin{equation}
        y=\frac{\Omega-\frac{\beta}{\sigma_b^2}}{\mathrm{ln(I_S(\Omega-\frac{2\beta}{\sigma_b^2}))-\mathrm{ln(I_{AS}(-\Omega))}}}-k_BT.
\end{equation}
The equation $y=0$ was solved numerically for $T$ at each energy $\Omega$ and then averaged over a finite frequency range at each time delay to improve the statistics. The accuracy of the temperature determination using this method is essentially limited by the weakness of the anti-Stokes Raman signal. At a large anti-Stokes shift, the signal rapidly becomes indistinguishable from the dark background of the CCD detector, whose actual value in the spectral region of interest is then the leading source of error. This is particularly problematic at low temperature, i.e., for negative time-delays, and below -150~cm$^{-1}$. For this reason, we chose in the manuscript to extract the temperature using the spectral range between 100 and 150 $\mathrm{cm}^{-1}$ where we have enough anti-Stokes signal even at negative delays and after 3~ps to extract a reliable temperature. In figure \ref{fig7} we compare the temperature obtained in the 100-150~cm$^{-1}$ and the 150-200~$\mathrm{cm}^{-1}$ range for delays between -1 and 3~ps using the above method. For negative delays and using our numerical algorithm, we could only extract a few data points with significant error bars for the 150-200~$\mathrm{cm}^{-1}$ spectral range due to the weakness of the anti-Stokes signal. Still, we observe essentially the same QP temperature dynamics for both spectral ranges with a slightly higher temperature of about 10~K between 0 and 1~ps at higher energy. After 1 ps, both datasets converge within the error bars.  


\begin{thebibliography}{50}%
\makeatletter
\providecommand \@ifxundefined [1]{%
 \@ifx{#1\undefined}
}%
\providecommand \@ifnum [1]{%
 \ifnum #1\expandafter \@firstoftwo
 \else \expandafter \@secondoftwo
 \fi
}%
\providecommand \@ifx [1]{%
 \ifx #1\expandafter \@firstoftwo
 \else \expandafter \@secondoftwo
 \fi
}%
\providecommand \natexlab [1]{#1}%
\providecommand \enquote  [1]{``#1''}%
\providecommand \bibnamefont  [1]{#1}%
\providecommand \bibfnamefont [1]{#1}%
\providecommand \citenamefont [1]{#1}%
\providecommand \href@noop [0]{\@secondoftwo}%
\providecommand \href [0]{\begingroup \@sanitize@url \@href}%
\providecommand \@href[1]{\@@startlink{#1}\@@href}%
\providecommand \@@href[1]{\endgroup#1\@@endlink}%
\providecommand \@sanitize@url [0]{\catcode `\\12\catcode `\$12\catcode `\&12\catcode `\#12\catcode `\^12\catcode `\_12\catcode `\%12\relax}%
\providecommand \@@startlink[1]{}%
\providecommand \@@endlink[0]{}%
\providecommand \url  [0]{\begingroup\@sanitize@url \@url }%
\providecommand \@url [1]{\endgroup\@href {#1}{\urlprefix }}%
\providecommand \urlprefix  [0]{URL }%
\providecommand \Eprint [0]{\href }%
\providecommand \doibase [0]{https://doi.org/}%
\providecommand \selectlanguage [0]{\@gobble}%
\providecommand \bibinfo  [0]{\@secondoftwo}%
\providecommand \bibfield  [0]{\@secondoftwo}%
\providecommand \translation [1]{[#1]}%
\providecommand \BibitemOpen [0]{}%
\providecommand \bibitemStop [0]{}%
\providecommand \bibitemNoStop [0]{.\EOS\space}%
\providecommand \EOS [0]{\spacefactor3000\relax}%
\providecommand \BibitemShut  [1]{\csname bibitem#1\endcsname}%
\let\auto@bib@innerbib\@empty
%</preamble>
\bibitem [{\citenamefont {Yonemitsu}\ and\ \citenamefont {Nasu}(2008)}]{yonemitsu_theory_2008}%
  \BibitemOpen
  \bibfield  {author} {\bibinfo {author} {\bibfnamefont {K.}~\bibnamefont {Yonemitsu}}\ and\ \bibinfo {author} {\bibfnamefont {K.}~\bibnamefont {Nasu}},\ }\href {https://doi.org/10.1016/j.physrep.2008.04.008} {\bibfield  {journal} {\bibinfo  {journal} {Physics Reports}\ }\textbf {\bibinfo {volume} {465}},\ \bibinfo {pages} {1} (\bibinfo {year} {2008})}\BibitemShut {NoStop}%
\bibitem [{\citenamefont {de~la Torre}\ \emph {et~al.}(2021)\citenamefont {de~la Torre}, \citenamefont {Kennes}, \citenamefont {Claassen}, \citenamefont {Gerber}, \citenamefont {McIver},\ and\ \citenamefont {Sentef}}]{de_la_torre_colloquium_2021}%
  \BibitemOpen
  \bibfield  {author} {\bibinfo {author} {\bibfnamefont {A.}~\bibnamefont {de~la Torre}}, \bibinfo {author} {\bibfnamefont {D.~M.}\ \bibnamefont {Kennes}}, \bibinfo {author} {\bibfnamefont {M.}~\bibnamefont {Claassen}}, \bibinfo {author} {\bibfnamefont {S.}~\bibnamefont {Gerber}}, \bibinfo {author} {\bibfnamefont {J.~W.}\ \bibnamefont {McIver}},\ and\ \bibinfo {author} {\bibfnamefont {M.~A.}\ \bibnamefont {Sentef}},\ }\href {https://doi.org/10.1103/RevModPhys.93.041002} {\bibfield  {journal} {\bibinfo  {journal} {Reviews of Modern Physics}\ }\textbf {\bibinfo {volume} {93}},\ \bibinfo {pages} {041002} (\bibinfo {year} {2021})}\BibitemShut {NoStop}%
\bibitem [{\citenamefont {Giannetti}\ \emph {et~al.}(2016)\citenamefont {Giannetti}, \citenamefont {Capone}, \citenamefont {Fausti}, \citenamefont {Fabrizio}, \citenamefont {Parmigiani},\ and\ \citenamefont {Mihailovic}}]{giannetti_ultrafast_2016}%
  \BibitemOpen
  \bibfield  {author} {\bibinfo {author} {\bibfnamefont {C.}~\bibnamefont {Giannetti}}, \bibinfo {author} {\bibfnamefont {M.}~\bibnamefont {Capone}}, \bibinfo {author} {\bibfnamefont {D.}~\bibnamefont {Fausti}}, \bibinfo {author} {\bibfnamefont {M.}~\bibnamefont {Fabrizio}}, \bibinfo {author} {\bibfnamefont {F.}~\bibnamefont {Parmigiani}},\ and\ \bibinfo {author} {\bibfnamefont {D.}~\bibnamefont {Mihailovic}},\ }\href {https://doi.org/10.1080/00018732.2016.1194044} {\bibfield  {journal} {\bibinfo  {journal} {Advances in Physics}\ }\textbf {\bibinfo {volume} {65}},\ \bibinfo {pages} {58} (\bibinfo {year} {2016})}\BibitemShut {NoStop}%
\bibitem [{\citenamefont {Pashkin}\ \emph {et~al.}(2010)\citenamefont {Pashkin}, \citenamefont {Porer}, \citenamefont {Beyer}, \citenamefont {Kim}, \citenamefont {Dubroka}, \citenamefont {Bernhard}, \citenamefont {Yao}, \citenamefont {Dagan}, \citenamefont {Hackl}, \citenamefont {Erb}, \citenamefont {Demsar}, \citenamefont {Huber},\ and\ \citenamefont {Leitenstorfer}}]{pashkin_femtosecond_2010}%
  \BibitemOpen
  \bibfield  {author} {\bibinfo {author} {\bibfnamefont {A.}~\bibnamefont {Pashkin}}, \bibinfo {author} {\bibfnamefont {M.}~\bibnamefont {Porer}}, \bibinfo {author} {\bibfnamefont {M.}~\bibnamefont {Beyer}}, \bibinfo {author} {\bibfnamefont {K.~W.}\ \bibnamefont {Kim}}, \bibinfo {author} {\bibfnamefont {A.}~\bibnamefont {Dubroka}}, \bibinfo {author} {\bibfnamefont {C.}~\bibnamefont {Bernhard}}, \bibinfo {author} {\bibfnamefont {X.}~\bibnamefont {Yao}}, \bibinfo {author} {\bibfnamefont {Y.}~\bibnamefont {Dagan}}, \bibinfo {author} {\bibfnamefont {R.}~\bibnamefont {Hackl}}, \bibinfo {author} {\bibfnamefont {A.}~\bibnamefont {Erb}}, \bibinfo {author} {\bibfnamefont {J.}~\bibnamefont {Demsar}}, \bibinfo {author} {\bibfnamefont {R.}~\bibnamefont {Huber}},\ and\ \bibinfo {author} {\bibfnamefont {A.}~\bibnamefont {Leitenstorfer}},\ }\href@noop {} {\bibfield  {journal} {\bibinfo  {journal} {Physical Review Letters}\ }\textbf {\bibinfo {volume} {105}},\ \bibinfo {pages} {067001} (\bibinfo {year} {2010})}\BibitemShut
  {NoStop}%
\bibitem [{\citenamefont {Porer}\ \emph {et~al.}(2014)\citenamefont {Porer}, \citenamefont {Leierseder}, \citenamefont {Ménard}, \citenamefont {Dachraoui}, \citenamefont {Mouchliadis}, \citenamefont {Perakis}, \citenamefont {Heinzmann}, \citenamefont {Demsar}, \citenamefont {Rossnagel},\ and\ \citenamefont {Huber}}]{porer_non-thermal_2014}%
  \BibitemOpen
  \bibfield  {author} {\bibinfo {author} {\bibfnamefont {M.}~\bibnamefont {Porer}}, \bibinfo {author} {\bibfnamefont {U.}~\bibnamefont {Leierseder}}, \bibinfo {author} {\bibfnamefont {J.-M.}\ \bibnamefont {Ménard}}, \bibinfo {author} {\bibfnamefont {H.}~\bibnamefont {Dachraoui}}, \bibinfo {author} {\bibfnamefont {L.}~\bibnamefont {Mouchliadis}}, \bibinfo {author} {\bibfnamefont {I.~E.}\ \bibnamefont {Perakis}}, \bibinfo {author} {\bibfnamefont {U.}~\bibnamefont {Heinzmann}}, \bibinfo {author} {\bibfnamefont {J.}~\bibnamefont {Demsar}}, \bibinfo {author} {\bibfnamefont {K.}~\bibnamefont {Rossnagel}},\ and\ \bibinfo {author} {\bibfnamefont {R.}~\bibnamefont {Huber}},\ }\href {https://doi.org/10.1038/nmat4042} {\bibfield  {journal} {\bibinfo  {journal} {Nature Materials}\ }\textbf {\bibinfo {volume} {13}},\ \bibinfo {pages} {857} (\bibinfo {year} {2014})}\BibitemShut {NoStop}%
\bibitem [{\citenamefont {Maklar}\ \emph {et~al.}(2021)\citenamefont {Maklar}, \citenamefont {Windsor}, \citenamefont {Nicholson}, \citenamefont {Puppin}, \citenamefont {Walmsley}, \citenamefont {Esposito}, \citenamefont {Porer}, \citenamefont {Rittmann}, \citenamefont {Leuenberger}, \citenamefont {Kubli}, \citenamefont {Savoini}, \citenamefont {Abreu}, \citenamefont {Johnson}, \citenamefont {Beaud}, \citenamefont {Ingold}, \citenamefont {Staub}, \citenamefont {Fisher}, \citenamefont {Ernstorfer}, \citenamefont {Wolf},\ and\ \citenamefont {Rettig}}]{maklar_nonequilibrium_2021}%
  \BibitemOpen
  \bibfield  {author} {\bibinfo {author} {\bibfnamefont {J.}~\bibnamefont {Maklar}}, \bibinfo {author} {\bibfnamefont {Y.~W.}\ \bibnamefont {Windsor}}, \bibinfo {author} {\bibfnamefont {C.~W.}\ \bibnamefont {Nicholson}}, \bibinfo {author} {\bibfnamefont {M.}~\bibnamefont {Puppin}}, \bibinfo {author} {\bibfnamefont {P.}~\bibnamefont {Walmsley}}, \bibinfo {author} {\bibfnamefont {V.}~\bibnamefont {Esposito}}, \bibinfo {author} {\bibfnamefont {M.}~\bibnamefont {Porer}}, \bibinfo {author} {\bibfnamefont {J.}~\bibnamefont {Rittmann}}, \bibinfo {author} {\bibfnamefont {D.}~\bibnamefont {Leuenberger}}, \bibinfo {author} {\bibfnamefont {M.}~\bibnamefont {Kubli}}, \bibinfo {author} {\bibfnamefont {M.}~\bibnamefont {Savoini}}, \bibinfo {author} {\bibfnamefont {E.}~\bibnamefont {Abreu}}, \bibinfo {author} {\bibfnamefont {S.~L.}\ \bibnamefont {Johnson}}, \bibinfo {author} {\bibfnamefont {P.}~\bibnamefont {Beaud}}, \bibinfo {author} {\bibfnamefont {G.}~\bibnamefont {Ingold}}, \bibinfo {author} {\bibfnamefont
  {U.}~\bibnamefont {Staub}}, \bibinfo {author} {\bibfnamefont {I.~R.}\ \bibnamefont {Fisher}}, \bibinfo {author} {\bibfnamefont {R.}~\bibnamefont {Ernstorfer}}, \bibinfo {author} {\bibfnamefont {M.}~\bibnamefont {Wolf}},\ and\ \bibinfo {author} {\bibfnamefont {L.}~\bibnamefont {Rettig}},\ }\href {https://doi.org/10.1038/s41467-021-22778-w} {\bibfield  {journal} {\bibinfo  {journal} {Nature Communications}\ }\textbf {\bibinfo {volume} {12}},\ \bibinfo {pages} {2499} (\bibinfo {year} {2021})}\BibitemShut {NoStop}%
\bibitem [{\citenamefont {Konstantinova}\ \emph {et~al.}(2018)\citenamefont {Konstantinova}, \citenamefont {Rameau}, \citenamefont {Reid}, \citenamefont {Abdurazakov}, \citenamefont {Wu}, \citenamefont {Li}, \citenamefont {Shen}, \citenamefont {Gu}, \citenamefont {Huang}, \citenamefont {Rettig}, \citenamefont {Avigo}, \citenamefont {Ligges}, \citenamefont {Freericks}, \citenamefont {Kemper}, \citenamefont {Dürr}, \citenamefont {Bovensiepen}, \citenamefont {Johnson}, \citenamefont {Wang},\ and\ \citenamefont {Zhu}}]{konstantinova_nonequilibrium_2018}%
  \BibitemOpen
  \bibfield  {author} {\bibinfo {author} {\bibfnamefont {T.}~\bibnamefont {Konstantinova}}, \bibinfo {author} {\bibfnamefont {J.~D.}\ \bibnamefont {Rameau}}, \bibinfo {author} {\bibfnamefont {A.~H.}\ \bibnamefont {Reid}}, \bibinfo {author} {\bibfnamefont {O.}~\bibnamefont {Abdurazakov}}, \bibinfo {author} {\bibfnamefont {L.}~\bibnamefont {Wu}}, \bibinfo {author} {\bibfnamefont {R.}~\bibnamefont {Li}}, \bibinfo {author} {\bibfnamefont {X.}~\bibnamefont {Shen}}, \bibinfo {author} {\bibfnamefont {G.}~\bibnamefont {Gu}}, \bibinfo {author} {\bibfnamefont {Y.}~\bibnamefont {Huang}}, \bibinfo {author} {\bibfnamefont {L.}~\bibnamefont {Rettig}}, \bibinfo {author} {\bibfnamefont {I.}~\bibnamefont {Avigo}}, \bibinfo {author} {\bibfnamefont {M.}~\bibnamefont {Ligges}}, \bibinfo {author} {\bibfnamefont {J.~K.}\ \bibnamefont {Freericks}}, \bibinfo {author} {\bibfnamefont {A.~F.}\ \bibnamefont {Kemper}}, \bibinfo {author} {\bibfnamefont {H.~A.}\ \bibnamefont {Dürr}}, \bibinfo {author} {\bibfnamefont {U.}~\bibnamefont
  {Bovensiepen}}, \bibinfo {author} {\bibfnamefont {P.~D.}\ \bibnamefont {Johnson}}, \bibinfo {author} {\bibfnamefont {X.}~\bibnamefont {Wang}},\ and\ \bibinfo {author} {\bibfnamefont {Y.}~\bibnamefont {Zhu}},\ }\href {https://doi.org/10.1126/sciadv.aap7427} {\bibfield  {journal} {\bibinfo  {journal} {Science Advances}\ }\textbf {\bibinfo {volume} {4}},\ \bibinfo {pages} {eaap7427} (\bibinfo {year} {2018})}\BibitemShut {NoStop}%
\bibitem [{\citenamefont {Parker}(1975)}]{parker_modified_1975}%
  \BibitemOpen
  \bibfield  {author} {\bibinfo {author} {\bibfnamefont {W.~H.}\ \bibnamefont {Parker}},\ }\href {https://doi.org/10.1103/PhysRevB.12.3667} {\bibfield  {journal} {\bibinfo  {journal} {Physical Review B}\ }\textbf {\bibinfo {volume} {12}},\ \bibinfo {pages} {3667} (\bibinfo {year} {1975})}\BibitemShut {NoStop}%
\bibitem [{\citenamefont {Allen}(1987)}]{allen_theory_1987}%
  \BibitemOpen
  \bibfield  {author} {\bibinfo {author} {\bibfnamefont {P.~B.}\ \bibnamefont {Allen}},\ }\href {https://doi.org/10.1103/PhysRevLett.59.1460} {\bibfield  {journal} {\bibinfo  {journal} {Physical Review Letters}\ }\textbf {\bibinfo {volume} {59}},\ \bibinfo {pages} {1460} (\bibinfo {year} {1987})}\BibitemShut {NoStop}%
\bibitem [{\citenamefont {Perfetti}\ \emph {et~al.}(2007)\citenamefont {Perfetti}, \citenamefont {Loukakos}, \citenamefont {Lisowski}, \citenamefont {Bovensiepen}, \citenamefont {Eisaki},\ and\ \citenamefont {Wolf}}]{perfetti_ultrafast_2007}%
  \BibitemOpen
  \bibfield  {author} {\bibinfo {author} {\bibfnamefont {L.}~\bibnamefont {Perfetti}}, \bibinfo {author} {\bibfnamefont {P.~A.}\ \bibnamefont {Loukakos}}, \bibinfo {author} {\bibfnamefont {M.}~\bibnamefont {Lisowski}}, \bibinfo {author} {\bibfnamefont {U.}~\bibnamefont {Bovensiepen}}, \bibinfo {author} {\bibfnamefont {H.}~\bibnamefont {Eisaki}},\ and\ \bibinfo {author} {\bibfnamefont {M.}~\bibnamefont {Wolf}},\ }\href {https://doi.org/10.1103/PhysRevLett.99.197001} {\bibfield  {journal} {\bibinfo  {journal} {Physical Review Letters}\ }\textbf {\bibinfo {volume} {99}},\ \bibinfo {pages} {197001} (\bibinfo {year} {2007})}\BibitemShut {NoStop}%
\bibitem [{\citenamefont {Demsar}(2020)}]{demsar_non-equilibrium_2020}%
  \BibitemOpen
  \bibfield  {author} {\bibinfo {author} {\bibfnamefont {J.}~\bibnamefont {Demsar}},\ }\href {https://doi.org/10.1007/s10909-020-02461-y} {\bibfield  {journal} {\bibinfo  {journal} {Journal of Low Temperature Physics}\ }\textbf {\bibinfo {volume} {201}},\ \bibinfo {pages} {676} (\bibinfo {year} {2020})}\BibitemShut {NoStop}%
\bibitem [{\citenamefont {Kusar}\ \emph {et~al.}(2008)\citenamefont {Kusar}, \citenamefont {Kabanov}, \citenamefont {Demsar}, \citenamefont {Mertelj}, \citenamefont {Sugai},\ and\ \citenamefont {Mihailovic}}]{kusar_controlled_2008}%
  \BibitemOpen
  \bibfield  {author} {\bibinfo {author} {\bibfnamefont {P.}~\bibnamefont {Kusar}}, \bibinfo {author} {\bibfnamefont {V.~V.}\ \bibnamefont {Kabanov}}, \bibinfo {author} {\bibfnamefont {J.}~\bibnamefont {Demsar}}, \bibinfo {author} {\bibfnamefont {T.}~\bibnamefont {Mertelj}}, \bibinfo {author} {\bibfnamefont {S.}~\bibnamefont {Sugai}},\ and\ \bibinfo {author} {\bibfnamefont {D.}~\bibnamefont {Mihailovic}},\ }\href {https://doi.org/10.1103/PhysRevLett.101.227001} {\bibfield  {journal} {\bibinfo  {journal} {Physical Review Letters}\ }\textbf {\bibinfo {volume} {101}},\ \bibinfo {pages} {227001} (\bibinfo {year} {2008})}\BibitemShut {NoStop}%
\bibitem [{\citenamefont {Giannetti}\ \emph {et~al.}(2009)\citenamefont {Giannetti}, \citenamefont {Coslovich}, \citenamefont {Cilento}, \citenamefont {Ferrini}, \citenamefont {Eisaki}, \citenamefont {Kaneko}, \citenamefont {Greven},\ and\ \citenamefont {Parmigiani}}]{giannetti_discontinuity_2009}%
  \BibitemOpen
  \bibfield  {author} {\bibinfo {author} {\bibfnamefont {C.}~\bibnamefont {Giannetti}}, \bibinfo {author} {\bibfnamefont {G.}~\bibnamefont {Coslovich}}, \bibinfo {author} {\bibfnamefont {F.}~\bibnamefont {Cilento}}, \bibinfo {author} {\bibfnamefont {G.}~\bibnamefont {Ferrini}}, \bibinfo {author} {\bibfnamefont {H.}~\bibnamefont {Eisaki}}, \bibinfo {author} {\bibfnamefont {N.}~\bibnamefont {Kaneko}}, \bibinfo {author} {\bibfnamefont {M.}~\bibnamefont {Greven}},\ and\ \bibinfo {author} {\bibfnamefont {F.}~\bibnamefont {Parmigiani}},\ }\href {https://doi.org/10.1103/PhysRevB.79.224502} {\bibfield  {journal} {\bibinfo  {journal} {Physical Review B}\ }\textbf {\bibinfo {volume} {79}},\ \bibinfo {pages} {224502} (\bibinfo {year} {2009})}\BibitemShut {NoStop}%
\bibitem [{\citenamefont {Coslovich}\ \emph {et~al.}(2011)\citenamefont {Coslovich}, \citenamefont {Giannetti}, \citenamefont {Cilento}, \citenamefont {Dal~Conte}, \citenamefont {Ferrini}, \citenamefont {Galinetto}, \citenamefont {Greven}, \citenamefont {Eisaki}, \citenamefont {Raichle}, \citenamefont {Liang}, \citenamefont {Damascelli},\ and\ \citenamefont {Parmigiani}}]{coslovich_evidence_2011}%
  \BibitemOpen
  \bibfield  {author} {\bibinfo {author} {\bibfnamefont {G.}~\bibnamefont {Coslovich}}, \bibinfo {author} {\bibfnamefont {C.}~\bibnamefont {Giannetti}}, \bibinfo {author} {\bibfnamefont {F.}~\bibnamefont {Cilento}}, \bibinfo {author} {\bibfnamefont {S.}~\bibnamefont {Dal~Conte}}, \bibinfo {author} {\bibfnamefont {G.}~\bibnamefont {Ferrini}}, \bibinfo {author} {\bibfnamefont {P.}~\bibnamefont {Galinetto}}, \bibinfo {author} {\bibfnamefont {M.}~\bibnamefont {Greven}}, \bibinfo {author} {\bibfnamefont {H.}~\bibnamefont {Eisaki}}, \bibinfo {author} {\bibfnamefont {M.}~\bibnamefont {Raichle}}, \bibinfo {author} {\bibfnamefont {R.}~\bibnamefont {Liang}}, \bibinfo {author} {\bibfnamefont {A.}~\bibnamefont {Damascelli}},\ and\ \bibinfo {author} {\bibfnamefont {F.}~\bibnamefont {Parmigiani}},\ }\href {https://doi.org/10.1103/PhysRevB.83.064519} {\bibfield  {journal} {\bibinfo  {journal} {Physical Review B}\ }\textbf {\bibinfo {volume} {83}},\ \bibinfo {pages} {064519} (\bibinfo {year} {2011})}\BibitemShut {NoStop}%
\bibitem [{\citenamefont {Madan}\ \emph {et~al.}(2017)\citenamefont {Madan}, \citenamefont {Baranov}, \citenamefont {Toda}, \citenamefont {Oda}, \citenamefont {Kurosawa}, \citenamefont {Kabanov}, \citenamefont {Mertelj},\ and\ \citenamefont {Mihailovic}}]{madan_dynamics_2017}%
  \BibitemOpen
  \bibfield  {author} {\bibinfo {author} {\bibfnamefont {I.}~\bibnamefont {Madan}}, \bibinfo {author} {\bibfnamefont {V.~V.}\ \bibnamefont {Baranov}}, \bibinfo {author} {\bibfnamefont {Y.}~\bibnamefont {Toda}}, \bibinfo {author} {\bibfnamefont {M.}~\bibnamefont {Oda}}, \bibinfo {author} {\bibfnamefont {T.}~\bibnamefont {Kurosawa}}, \bibinfo {author} {\bibfnamefont {V.~V.}\ \bibnamefont {Kabanov}}, \bibinfo {author} {\bibfnamefont {T.}~\bibnamefont {Mertelj}},\ and\ \bibinfo {author} {\bibfnamefont {D.}~\bibnamefont {Mihailovic}},\ }\href {https://doi.org/10.1103/PhysRevB.96.184522} {\bibfield  {journal} {\bibinfo  {journal} {Physical Review B}\ }\textbf {\bibinfo {volume} {96}},\ \bibinfo {pages} {184522} (\bibinfo {year} {2017})}\BibitemShut {NoStop}%
\bibitem [{\citenamefont {Carnahan}\ \emph {et~al.}(2004)\citenamefont {Carnahan}, \citenamefont {Kaindl}, \citenamefont {Orenstein}, \citenamefont {Chemla}, \citenamefont {Oh},\ and\ \citenamefont {Eckstein}}]{carnahan_nonequilibrium_2004}%
  \BibitemOpen
  \bibfield  {author} {\bibinfo {author} {\bibfnamefont {M.~A.}\ \bibnamefont {Carnahan}}, \bibinfo {author} {\bibfnamefont {R.~A.}\ \bibnamefont {Kaindl}}, \bibinfo {author} {\bibfnamefont {J.}~\bibnamefont {Orenstein}}, \bibinfo {author} {\bibfnamefont {D.~S.}\ \bibnamefont {Chemla}}, \bibinfo {author} {\bibfnamefont {S.}~\bibnamefont {Oh}},\ and\ \bibinfo {author} {\bibfnamefont {J.~N.}\ \bibnamefont {Eckstein}},\ }\href {https://doi.org/10.1016/j.physc.2004.03.056} {\bibfield  {journal} {\bibinfo  {journal} {Physica C: Superconductivity}\ }\bibinfo {series} {Proceedings of the {International} {Conference} on {Materials} and {Mechanisms} of {Superconductivity}. {High} {Temperature} {Superconductors} {VII} -- {M2SRIO}},\ \textbf {\bibinfo {volume} {408-410}},\ \bibinfo {pages} {729} (\bibinfo {year} {2004})}\BibitemShut {NoStop}%
\bibitem [{\citenamefont {Beyer}\ \emph {et~al.}(2011)\citenamefont {Beyer}, \citenamefont {Städter}, \citenamefont {Beck}, \citenamefont {Schäfer}, \citenamefont {Kabanov}, \citenamefont {Logvenov}, \citenamefont {Bozovic}, \citenamefont {Koren},\ and\ \citenamefont {Demsar}}]{beyer_photoinduced_2011}%
  \BibitemOpen
  \bibfield  {author} {\bibinfo {author} {\bibfnamefont {M.}~\bibnamefont {Beyer}}, \bibinfo {author} {\bibfnamefont {D.}~\bibnamefont {Städter}}, \bibinfo {author} {\bibfnamefont {M.}~\bibnamefont {Beck}}, \bibinfo {author} {\bibfnamefont {H.}~\bibnamefont {Schäfer}}, \bibinfo {author} {\bibfnamefont {V.~V.}\ \bibnamefont {Kabanov}}, \bibinfo {author} {\bibfnamefont {G.}~\bibnamefont {Logvenov}}, \bibinfo {author} {\bibfnamefont {I.}~\bibnamefont {Bozovic}}, \bibinfo {author} {\bibfnamefont {G.}~\bibnamefont {Koren}},\ and\ \bibinfo {author} {\bibfnamefont {J.}~\bibnamefont {Demsar}},\ }\href {https://doi.org/10.1103/PhysRevB.83.214515} {\bibfield  {journal} {\bibinfo  {journal} {Physical Review B}\ }\textbf {\bibinfo {volume} {83}},\ \bibinfo {pages} {214515} (\bibinfo {year} {2011})}\BibitemShut {NoStop}%
\bibitem [{\citenamefont {Beck}\ \emph {et~al.}(2017)\citenamefont {Beck}, \citenamefont {Klammer}, \citenamefont {Rousseau}, \citenamefont {Obergfell}, \citenamefont {Leiderer}, \citenamefont {Helm}, \citenamefont {Kabanov}, \citenamefont {Diamant}, \citenamefont {Rabinowicz}, \citenamefont {Dagan},\ and\ \citenamefont {Demsar}}]{beck_energy_2017}%
  \BibitemOpen
  \bibfield  {author} {\bibinfo {author} {\bibfnamefont {M.}~\bibnamefont {Beck}}, \bibinfo {author} {\bibfnamefont {M.}~\bibnamefont {Klammer}}, \bibinfo {author} {\bibfnamefont {I.}~\bibnamefont {Rousseau}}, \bibinfo {author} {\bibfnamefont {M.}~\bibnamefont {Obergfell}}, \bibinfo {author} {\bibfnamefont {P.}~\bibnamefont {Leiderer}}, \bibinfo {author} {\bibfnamefont {M.}~\bibnamefont {Helm}}, \bibinfo {author} {\bibfnamefont {V.~V.}\ \bibnamefont {Kabanov}}, \bibinfo {author} {\bibfnamefont {I.}~\bibnamefont {Diamant}}, \bibinfo {author} {\bibfnamefont {A.}~\bibnamefont {Rabinowicz}}, \bibinfo {author} {\bibfnamefont {Y.}~\bibnamefont {Dagan}},\ and\ \bibinfo {author} {\bibfnamefont {J.}~\bibnamefont {Demsar}},\ }\href {https://doi.org/10.1103/PhysRevB.95.085106} {\bibfield  {journal} {\bibinfo  {journal} {Physical Review B}\ }\textbf {\bibinfo {volume} {95}},\ \bibinfo {pages} {085106} (\bibinfo {year} {2017})}\BibitemShut {NoStop}%
\bibitem [{\citenamefont {Cortés}\ \emph {et~al.}(2011)\citenamefont {Cortés}, \citenamefont {Rettig}, \citenamefont {Yoshida}, \citenamefont {Eisaki}, \citenamefont {Wolf},\ and\ \citenamefont {Bovensiepen}}]{cortes_momentum-resolved_2011}%
  \BibitemOpen
  \bibfield  {author} {\bibinfo {author} {\bibfnamefont {R.}~\bibnamefont {Cortés}}, \bibinfo {author} {\bibfnamefont {L.}~\bibnamefont {Rettig}}, \bibinfo {author} {\bibfnamefont {Y.}~\bibnamefont {Yoshida}}, \bibinfo {author} {\bibfnamefont {H.}~\bibnamefont {Eisaki}}, \bibinfo {author} {\bibfnamefont {M.}~\bibnamefont {Wolf}},\ and\ \bibinfo {author} {\bibfnamefont {U.}~\bibnamefont {Bovensiepen}},\ }\href {https://doi.org/10.1103/PhysRevLett.107.097002} {\bibfield  {journal} {\bibinfo  {journal} {Physical Review Letters}\ }\textbf {\bibinfo {volume} {107}},\ \bibinfo {pages} {097002} (\bibinfo {year} {2011})}\BibitemShut {NoStop}%
\bibitem [{\citenamefont {Smallwood}\ \emph {et~al.}(2012)\citenamefont {Smallwood}, \citenamefont {Hinton}, \citenamefont {Jozwiak}, \citenamefont {Zhang}, \citenamefont {Koralek}, \citenamefont {Eisaki}, \citenamefont {Lee}, \citenamefont {Orenstein},\ and\ \citenamefont {Lanzara}}]{smallwood_tracking_2012}%
  \BibitemOpen
  \bibfield  {author} {\bibinfo {author} {\bibfnamefont {C.~L.}\ \bibnamefont {Smallwood}}, \bibinfo {author} {\bibfnamefont {J.~P.}\ \bibnamefont {Hinton}}, \bibinfo {author} {\bibfnamefont {C.}~\bibnamefont {Jozwiak}}, \bibinfo {author} {\bibfnamefont {W.}~\bibnamefont {Zhang}}, \bibinfo {author} {\bibfnamefont {J.~D.}\ \bibnamefont {Koralek}}, \bibinfo {author} {\bibfnamefont {H.}~\bibnamefont {Eisaki}}, \bibinfo {author} {\bibfnamefont {D.-H.}\ \bibnamefont {Lee}}, \bibinfo {author} {\bibfnamefont {J.}~\bibnamefont {Orenstein}},\ and\ \bibinfo {author} {\bibfnamefont {A.}~\bibnamefont {Lanzara}},\ }\href {https://doi.org/10.1126/science.1217423} {\bibfield  {journal} {\bibinfo  {journal} {Science}\ }\textbf {\bibinfo {volume} {336}},\ \bibinfo {pages} {1137} (\bibinfo {year} {2012})}\BibitemShut {NoStop}%
\bibitem [{\citenamefont {Smallwood}\ \emph {et~al.}(2016)\citenamefont {Smallwood}, \citenamefont {Miller}, \citenamefont {Zhang}, \citenamefont {Kaindl},\ and\ \citenamefont {Lanzara}}]{smallwood_nonequilibrium_2016}%
  \BibitemOpen
  \bibfield  {author} {\bibinfo {author} {\bibfnamefont {C.~L.}\ \bibnamefont {Smallwood}}, \bibinfo {author} {\bibfnamefont {T.~L.}\ \bibnamefont {Miller}}, \bibinfo {author} {\bibfnamefont {W.}~\bibnamefont {Zhang}}, \bibinfo {author} {\bibfnamefont {R.~A.}\ \bibnamefont {Kaindl}},\ and\ \bibinfo {author} {\bibfnamefont {A.}~\bibnamefont {Lanzara}},\ }\href {https://doi.org/10.1103/PhysRevB.93.235107} {\bibfield  {journal} {\bibinfo  {journal} {Physical Review B}\ }\textbf {\bibinfo {volume} {93}},\ \bibinfo {pages} {235107} (\bibinfo {year} {2016})}\BibitemShut {NoStop}%
\bibitem [{\citenamefont {Miller}\ \emph {et~al.}(2015)\citenamefont {Miller}, \citenamefont {Smallwood}, \citenamefont {Zhang}, \citenamefont {Eisaki}, \citenamefont {Orenstein},\ and\ \citenamefont {Lanzara}}]{miller_photoinduced_2015}%
  \BibitemOpen
  \bibfield  {author} {\bibinfo {author} {\bibfnamefont {T.~L.}\ \bibnamefont {Miller}}, \bibinfo {author} {\bibfnamefont {C.~L.}\ \bibnamefont {Smallwood}}, \bibinfo {author} {\bibfnamefont {W.}~\bibnamefont {Zhang}}, \bibinfo {author} {\bibfnamefont {H.}~\bibnamefont {Eisaki}}, \bibinfo {author} {\bibfnamefont {J.}~\bibnamefont {Orenstein}},\ and\ \bibinfo {author} {\bibfnamefont {A.}~\bibnamefont {Lanzara}},\ }\href {https://doi.org/10.1103/PhysRevB.92.144506} {\bibfield  {journal} {\bibinfo  {journal} {Physical Review B}\ }\textbf {\bibinfo {volume} {92}},\ \bibinfo {pages} {144506} (\bibinfo {year} {2015})}\BibitemShut {NoStop}%
\bibitem [{\citenamefont {Piovera}\ \emph {et~al.}(2015)\citenamefont {Piovera}, \citenamefont {Zhang}, \citenamefont {d'Astuto}, \citenamefont {Taleb-Ibrahimi}, \citenamefont {Papalazarou}, \citenamefont {Marsi}, \citenamefont {Li}, \citenamefont {Raffy},\ and\ \citenamefont {Perfetti}}]{piovera_quasiparticle_2015}%
  \BibitemOpen
  \bibfield  {author} {\bibinfo {author} {\bibfnamefont {C.}~\bibnamefont {Piovera}}, \bibinfo {author} {\bibfnamefont {Z.}~\bibnamefont {Zhang}}, \bibinfo {author} {\bibfnamefont {M.}~\bibnamefont {d'Astuto}}, \bibinfo {author} {\bibfnamefont {A.}~\bibnamefont {Taleb-Ibrahimi}}, \bibinfo {author} {\bibfnamefont {E.}~\bibnamefont {Papalazarou}}, \bibinfo {author} {\bibfnamefont {M.}~\bibnamefont {Marsi}}, \bibinfo {author} {\bibfnamefont {Z.~Z.}\ \bibnamefont {Li}}, \bibinfo {author} {\bibfnamefont {H.}~\bibnamefont {Raffy}},\ and\ \bibinfo {author} {\bibfnamefont {L.}~\bibnamefont {Perfetti}},\ }\href {https://doi.org/10.1103/PhysRevB.91.224509} {\bibfield  {journal} {\bibinfo  {journal} {Physical Review B}\ }\textbf {\bibinfo {volume} {91}},\ \bibinfo {pages} {224509} (\bibinfo {year} {2015})}\BibitemShut {NoStop}%
\bibitem [{\citenamefont {Ishida}\ \emph {et~al.}(2016)\citenamefont {Ishida}, \citenamefont {Saitoh}, \citenamefont {Mochiku}, \citenamefont {Nakane}, \citenamefont {Hirata},\ and\ \citenamefont {Shin}}]{ishida_quasi-particles_2016}%
  \BibitemOpen
  \bibfield  {author} {\bibinfo {author} {\bibfnamefont {Y.}~\bibnamefont {Ishida}}, \bibinfo {author} {\bibfnamefont {T.}~\bibnamefont {Saitoh}}, \bibinfo {author} {\bibfnamefont {T.}~\bibnamefont {Mochiku}}, \bibinfo {author} {\bibfnamefont {T.}~\bibnamefont {Nakane}}, \bibinfo {author} {\bibfnamefont {K.}~\bibnamefont {Hirata}},\ and\ \bibinfo {author} {\bibfnamefont {S.}~\bibnamefont {Shin}},\ }\href {https://doi.org/10.1038/srep18747} {\bibfield  {journal} {\bibinfo  {journal} {Scientific Reports}\ }\textbf {\bibinfo {volume} {6}},\ \bibinfo {pages} {18747} (\bibinfo {year} {2016})}\BibitemShut {NoStop}%
\bibitem [{\citenamefont {Zhang}\ \emph {et~al.}(2017)\citenamefont {Zhang}, \citenamefont {Piovera}, \citenamefont {Papalazarou}, \citenamefont {Marsi}, \citenamefont {d'Astuto}, \citenamefont {van~der Beek}, \citenamefont {Taleb-Ibrahimi},\ and\ \citenamefont {Perfetti}}]{zhang_photoinduced_2017}%
  \BibitemOpen
  \bibfield  {author} {\bibinfo {author} {\bibfnamefont {Z.}~\bibnamefont {Zhang}}, \bibinfo {author} {\bibfnamefont {C.}~\bibnamefont {Piovera}}, \bibinfo {author} {\bibfnamefont {E.}~\bibnamefont {Papalazarou}}, \bibinfo {author} {\bibfnamefont {M.}~\bibnamefont {Marsi}}, \bibinfo {author} {\bibfnamefont {M.}~\bibnamefont {d'Astuto}}, \bibinfo {author} {\bibfnamefont {C.~J.}\ \bibnamefont {van~der Beek}}, \bibinfo {author} {\bibfnamefont {A.}~\bibnamefont {Taleb-Ibrahimi}},\ and\ \bibinfo {author} {\bibfnamefont {L.}~\bibnamefont {Perfetti}},\ }\href {https://doi.org/10.1103/PhysRevB.96.064510} {\bibfield  {journal} {\bibinfo  {journal} {Physical Review B}\ }\textbf {\bibinfo {volume} {96}},\ \bibinfo {pages} {064510} (\bibinfo {year} {2017})}\BibitemShut {NoStop}%
\bibitem [{\citenamefont {Parham}\ \emph {et~al.}(2017)\citenamefont {Parham}, \citenamefont {Li}, \citenamefont {Nummy}, \citenamefont {Waugh}, \citenamefont {Zhou}, \citenamefont {Griffith}, \citenamefont {Schneeloch}, \citenamefont {Zhong}, \citenamefont {Gu},\ and\ \citenamefont {Dessau}}]{parham_ultrafast_2017}%
  \BibitemOpen
  \bibfield  {author} {\bibinfo {author} {\bibfnamefont {S.}~\bibnamefont {Parham}}, \bibinfo {author} {\bibfnamefont {H.}~\bibnamefont {Li}}, \bibinfo {author} {\bibfnamefont {T.}~\bibnamefont {Nummy}}, \bibinfo {author} {\bibfnamefont {J.}~\bibnamefont {Waugh}}, \bibinfo {author} {\bibfnamefont {X.}~\bibnamefont {Zhou}}, \bibinfo {author} {\bibfnamefont {J.}~\bibnamefont {Griffith}}, \bibinfo {author} {\bibfnamefont {J.}~\bibnamefont {Schneeloch}}, \bibinfo {author} {\bibfnamefont {R.}~\bibnamefont {Zhong}}, \bibinfo {author} {\bibfnamefont {G.}~\bibnamefont {Gu}},\ and\ \bibinfo {author} {\bibfnamefont {D.}~\bibnamefont {Dessau}},\ }\href {https://doi.org/10.1103/PhysRevX.7.041013} {\bibfield  {journal} {\bibinfo  {journal} {Physical Review X}\ }\textbf {\bibinfo {volume} {7}},\ \bibinfo {pages} {041013} (\bibinfo {year} {2017})}\BibitemShut {NoStop}%
\bibitem [{\citenamefont {Boschini}\ \emph {et~al.}(2018)\citenamefont {Boschini}, \citenamefont {da~Silva~Neto}, \citenamefont {Razzoli}, \citenamefont {Zonno}, \citenamefont {Peli}, \citenamefont {Day}, \citenamefont {Michiardi}, \citenamefont {Schneider}, \citenamefont {Zwartsenberg}, \citenamefont {Nigge}, \citenamefont {Zhong}, \citenamefont {Schneeloch}, \citenamefont {Gu}, \citenamefont {Zhdanovich}, \citenamefont {Mills}, \citenamefont {Levy}, \citenamefont {Jones}, \citenamefont {Giannetti},\ and\ \citenamefont {Damascelli}}]{boschini_collapse_2018}%
  \BibitemOpen
  \bibfield  {author} {\bibinfo {author} {\bibfnamefont {F.}~\bibnamefont {Boschini}}, \bibinfo {author} {\bibfnamefont {E.~H.}\ \bibnamefont {da~Silva~Neto}}, \bibinfo {author} {\bibfnamefont {E.}~\bibnamefont {Razzoli}}, \bibinfo {author} {\bibfnamefont {M.}~\bibnamefont {Zonno}}, \bibinfo {author} {\bibfnamefont {S.}~\bibnamefont {Peli}}, \bibinfo {author} {\bibfnamefont {R.~P.}\ \bibnamefont {Day}}, \bibinfo {author} {\bibfnamefont {M.}~\bibnamefont {Michiardi}}, \bibinfo {author} {\bibfnamefont {M.}~\bibnamefont {Schneider}}, \bibinfo {author} {\bibfnamefont {B.}~\bibnamefont {Zwartsenberg}}, \bibinfo {author} {\bibfnamefont {P.}~\bibnamefont {Nigge}}, \bibinfo {author} {\bibfnamefont {R.~D.}\ \bibnamefont {Zhong}}, \bibinfo {author} {\bibfnamefont {J.}~\bibnamefont {Schneeloch}}, \bibinfo {author} {\bibfnamefont {G.~D.}\ \bibnamefont {Gu}}, \bibinfo {author} {\bibfnamefont {S.}~\bibnamefont {Zhdanovich}}, \bibinfo {author} {\bibfnamefont {A.~K.}\ \bibnamefont {Mills}}, \bibinfo {author} {\bibfnamefont
  {G.}~\bibnamefont {Levy}}, \bibinfo {author} {\bibfnamefont {D.~J.}\ \bibnamefont {Jones}}, \bibinfo {author} {\bibfnamefont {C.}~\bibnamefont {Giannetti}},\ and\ \bibinfo {author} {\bibfnamefont {A.}~\bibnamefont {Damascelli}},\ }\href {https://doi.org/10.1038/s41563-018-0045-1} {\bibfield  {journal} {\bibinfo  {journal} {Nature Materials}\ }\textbf {\bibinfo {volume} {17}},\ \bibinfo {pages} {416} (\bibinfo {year} {2018})}\BibitemShut {NoStop}%
\bibitem [{\citenamefont {Zonno}\ \emph {et~al.}(2021)\citenamefont {Zonno}, \citenamefont {Boschini},\ and\ \citenamefont {Damascelli}}]{zonno_time-resolved_2021}%
  \BibitemOpen
  \bibfield  {author} {\bibinfo {author} {\bibfnamefont {M.}~\bibnamefont {Zonno}}, \bibinfo {author} {\bibfnamefont {F.}~\bibnamefont {Boschini}},\ and\ \bibinfo {author} {\bibfnamefont {A.}~\bibnamefont {Damascelli}},\ }\href {https://doi.org/10.1016/j.elspec.2021.147091} {\bibfield  {journal} {\bibinfo  {journal} {Journal of Electron Spectroscopy and Related Phenomena}\ }\textbf {\bibinfo {volume} {251}},\ \bibinfo {pages} {147091} (\bibinfo {year} {2021})}\BibitemShut {NoStop}%
\bibitem [{\citenamefont {Armanno}\ \emph {et~al.}(2025)\citenamefont {Armanno}, \citenamefont {Goto}, \citenamefont {Parent}, \citenamefont {Lapointe}, \citenamefont {Longa}, \citenamefont {Zhong}, \citenamefont {Schneeloch}, \citenamefont {Gu}, \citenamefont {Jargot}, \citenamefont {Ibrahim}, \citenamefont {Legare}, \citenamefont {Siwick}, \citenamefont {Gauthier},\ and\ \citenamefont {Boschini}}]{armanno_direct_2025}%
  \BibitemOpen
  \bibfield  {author} {\bibinfo {author} {\bibfnamefont {D.}~\bibnamefont {Armanno}}, \bibinfo {author} {\bibfnamefont {F.}~\bibnamefont {Goto}}, \bibinfo {author} {\bibfnamefont {J.-M.}\ \bibnamefont {Parent}}, \bibinfo {author} {\bibfnamefont {S.}~\bibnamefont {Lapointe}}, \bibinfo {author} {\bibfnamefont {A.}~\bibnamefont {Longa}}, \bibinfo {author} {\bibfnamefont {R.~D.}\ \bibnamefont {Zhong}}, \bibinfo {author} {\bibfnamefont {J.}~\bibnamefont {Schneeloch}}, \bibinfo {author} {\bibfnamefont {G.~D.}\ \bibnamefont {Gu}}, \bibinfo {author} {\bibfnamefont {G.}~\bibnamefont {Jargot}}, \bibinfo {author} {\bibfnamefont {H.}~\bibnamefont {Ibrahim}}, \bibinfo {author} {\bibfnamefont {F.}~\bibnamefont {Legare}}, \bibinfo {author} {\bibfnamefont {B.~J.}\ \bibnamefont {Siwick}}, \bibinfo {author} {\bibfnamefont {N.}~\bibnamefont {Gauthier}},\ and\ \bibinfo {author} {\bibfnamefont {F.}~\bibnamefont {Boschini}},\ }\href {https://doi.org/10.48550/arXiv.2505.03900} {\bibinfo {title} {Direct evidence of light-induced
  phase-fluctuations in cuprates via time-resolved {ARPES}}} (\bibinfo {year} {2025}),\ \bibinfo {note} {arXiv:2505.03900 [cond-mat]}\BibitemShut {NoStop}%
\bibitem [{\citenamefont {Kondo}\ \emph {et~al.}(2015)\citenamefont {Kondo}, \citenamefont {Malaeb}, \citenamefont {Ishida}, \citenamefont {Sasagawa}, \citenamefont {Sakamoto}, \citenamefont {Takeuchi}, \citenamefont {Tohyama},\ and\ \citenamefont {Shin}}]{kondo_point_2015}%
  \BibitemOpen
  \bibfield  {author} {\bibinfo {author} {\bibfnamefont {T.}~\bibnamefont {Kondo}}, \bibinfo {author} {\bibfnamefont {W.}~\bibnamefont {Malaeb}}, \bibinfo {author} {\bibfnamefont {Y.}~\bibnamefont {Ishida}}, \bibinfo {author} {\bibfnamefont {T.}~\bibnamefont {Sasagawa}}, \bibinfo {author} {\bibfnamefont {H.}~\bibnamefont {Sakamoto}}, \bibinfo {author} {\bibfnamefont {T.}~\bibnamefont {Takeuchi}}, \bibinfo {author} {\bibfnamefont {T.}~\bibnamefont {Tohyama}},\ and\ \bibinfo {author} {\bibfnamefont {S.}~\bibnamefont {Shin}},\ }\href {https://doi.org/10.1038/ncomms8699} {\bibfield  {journal} {\bibinfo  {journal} {Nature Communications}\ }\textbf {\bibinfo {volume} {6}},\ \bibinfo {pages} {7699} (\bibinfo {year} {2015})}\BibitemShut {NoStop}%
\bibitem [{\citenamefont {Dakovski}\ \emph {et~al.}(2015)\citenamefont {Dakovski}, \citenamefont {Durakiewicz}, \citenamefont {Zhu}, \citenamefont {Riseborough}, \citenamefont {Gu}, \citenamefont {Gilbertson}, \citenamefont {Taylor},\ and\ \citenamefont {Rodriguez}}]{dakovski_quasiparticle_2015}%
  \BibitemOpen
  \bibfield  {author} {\bibinfo {author} {\bibfnamefont {G.~L.}\ \bibnamefont {Dakovski}}, \bibinfo {author} {\bibfnamefont {T.}~\bibnamefont {Durakiewicz}}, \bibinfo {author} {\bibfnamefont {J.-X.}\ \bibnamefont {Zhu}}, \bibinfo {author} {\bibfnamefont {P.~S.}\ \bibnamefont {Riseborough}}, \bibinfo {author} {\bibfnamefont {G.}~\bibnamefont {Gu}}, \bibinfo {author} {\bibfnamefont {S.~M.}\ \bibnamefont {Gilbertson}}, \bibinfo {author} {\bibfnamefont {A.}~\bibnamefont {Taylor}},\ and\ \bibinfo {author} {\bibfnamefont {G.}~\bibnamefont {Rodriguez}},\ }\href {https://doi.org/10.1063/1.4933133} {\bibfield  {journal} {\bibinfo  {journal} {Structural Dynamics}\ }\textbf {\bibinfo {volume} {2}},\ \bibinfo {pages} {054501} (\bibinfo {year} {2015})}\BibitemShut {NoStop}%
\bibitem [{\citenamefont {Cilento}\ \emph {et~al.}(2018)\citenamefont {Cilento}, \citenamefont {Manzoni}, \citenamefont {Sterzi}, \citenamefont {Peli}, \citenamefont {Ronchi}, \citenamefont {Crepaldi}, \citenamefont {Boschini}, \citenamefont {Cacho}, \citenamefont {Chapman}, \citenamefont {Springate}, \citenamefont {Eisaki}, \citenamefont {Greven}, \citenamefont {Berciu}, \citenamefont {Kemper}, \citenamefont {Damascelli}, \citenamefont {Capone}, \citenamefont {Giannetti},\ and\ \citenamefont {Parmigiani}}]{cilento_dynamics_2018}%
  \BibitemOpen
  \bibfield  {author} {\bibinfo {author} {\bibfnamefont {F.}~\bibnamefont {Cilento}}, \bibinfo {author} {\bibfnamefont {G.}~\bibnamefont {Manzoni}}, \bibinfo {author} {\bibfnamefont {A.}~\bibnamefont {Sterzi}}, \bibinfo {author} {\bibfnamefont {S.}~\bibnamefont {Peli}}, \bibinfo {author} {\bibfnamefont {A.}~\bibnamefont {Ronchi}}, \bibinfo {author} {\bibfnamefont {A.}~\bibnamefont {Crepaldi}}, \bibinfo {author} {\bibfnamefont {F.}~\bibnamefont {Boschini}}, \bibinfo {author} {\bibfnamefont {C.}~\bibnamefont {Cacho}}, \bibinfo {author} {\bibfnamefont {R.}~\bibnamefont {Chapman}}, \bibinfo {author} {\bibfnamefont {E.}~\bibnamefont {Springate}}, \bibinfo {author} {\bibfnamefont {H.}~\bibnamefont {Eisaki}}, \bibinfo {author} {\bibfnamefont {M.}~\bibnamefont {Greven}}, \bibinfo {author} {\bibfnamefont {M.}~\bibnamefont {Berciu}}, \bibinfo {author} {\bibfnamefont {A.~F.}\ \bibnamefont {Kemper}}, \bibinfo {author} {\bibfnamefont {A.}~\bibnamefont {Damascelli}}, \bibinfo {author} {\bibfnamefont {M.}~\bibnamefont
  {Capone}}, \bibinfo {author} {\bibfnamefont {C.}~\bibnamefont {Giannetti}},\ and\ \bibinfo {author} {\bibfnamefont {F.}~\bibnamefont {Parmigiani}},\ }\href {https://doi.org/10.1126/sciadv.aar1998} {\bibfield  {journal} {\bibinfo  {journal} {Science Advances}\ }\textbf {\bibinfo {volume} {4}},\ \bibinfo {pages} {eaar1998} (\bibinfo {year} {2018})}\BibitemShut {NoStop}%
\bibitem [{\citenamefont {{H. Molegraaf}}(2005)}]{molegraaf_signposts_2005}%
  \BibitemOpen
  \bibfield  {author} {\bibinfo {author} {\bibnamefont {{H. Molegraaf}}},\ }\emph {\bibinfo {title} {Signposts to the mechanism of superconductivity in the cuprates}},\ \href@noop {} {Ph.D. thesis},\ \bibinfo  {school} {Rijksuniversiteit Groningen} (\bibinfo {year} {2005})\BibitemShut {NoStop}%
\bibitem [{\citenamefont {Versteeg}\ \emph {et~al.}(2018)\citenamefont {Versteeg}, \citenamefont {Zhu}, \citenamefont {Padmanabhan}, \citenamefont {Boguschewski}, \citenamefont {German}, \citenamefont {Goedecke}, \citenamefont {Becker},\ and\ \citenamefont {van Loosdrecht}}]{versteeg_tunable_2018}%
  \BibitemOpen
  \bibfield  {author} {\bibinfo {author} {\bibfnamefont {R.~B.}\ \bibnamefont {Versteeg}}, \bibinfo {author} {\bibfnamefont {J.}~\bibnamefont {Zhu}}, \bibinfo {author} {\bibfnamefont {P.}~\bibnamefont {Padmanabhan}}, \bibinfo {author} {\bibfnamefont {C.}~\bibnamefont {Boguschewski}}, \bibinfo {author} {\bibfnamefont {R.}~\bibnamefont {German}}, \bibinfo {author} {\bibfnamefont {M.}~\bibnamefont {Goedecke}}, \bibinfo {author} {\bibfnamefont {P.}~\bibnamefont {Becker}},\ and\ \bibinfo {author} {\bibfnamefont {P.~H.~M.}\ \bibnamefont {van Loosdrecht}},\ }\href {https://doi.org/10.1063/1.5037784} {\bibfield  {journal} {\bibinfo  {journal} {Structural Dynamics}\ }\textbf {\bibinfo {volume} {5}},\ \bibinfo {pages} {044301} (\bibinfo {year} {2018})}\BibitemShut {NoStop}%
\bibitem [{\citenamefont {Devereaux}\ and\ \citenamefont {Hackl}(2007)}]{devereaux_inelastic_2007}%
  \BibitemOpen
  \bibfield  {author} {\bibinfo {author} {\bibfnamefont {T.~P.}\ \bibnamefont {Devereaux}}\ and\ \bibinfo {author} {\bibfnamefont {R.}~\bibnamefont {Hackl}},\ }\href {https://doi.org/10.1103/RevModPhys.79.175} {\bibfield  {journal} {\bibinfo  {journal} {Reviews of Modern Physics}\ }\textbf {\bibinfo {volume} {79}},\ \bibinfo {pages} {175} (\bibinfo {year} {2007})}\BibitemShut {NoStop}%
\bibitem [{\citenamefont {Blanc}\ \emph {et~al.}(2010)\citenamefont {Blanc}, \citenamefont {Gallais}, \citenamefont {Cazayous}, \citenamefont {Méasson}, \citenamefont {Sacuto}, \citenamefont {Georges}, \citenamefont {Wen}, \citenamefont {Xu}, \citenamefont {Gu},\ and\ \citenamefont {Colson}}]{blanc_loss_2010}%
  \BibitemOpen
  \bibfield  {author} {\bibinfo {author} {\bibfnamefont {S.}~\bibnamefont {Blanc}}, \bibinfo {author} {\bibfnamefont {Y.}~\bibnamefont {Gallais}}, \bibinfo {author} {\bibfnamefont {M.}~\bibnamefont {Cazayous}}, \bibinfo {author} {\bibfnamefont {M.~A.}\ \bibnamefont {Méasson}}, \bibinfo {author} {\bibfnamefont {A.}~\bibnamefont {Sacuto}}, \bibinfo {author} {\bibfnamefont {A.}~\bibnamefont {Georges}}, \bibinfo {author} {\bibfnamefont {J.~S.}\ \bibnamefont {Wen}}, \bibinfo {author} {\bibfnamefont {Z.~J.}\ \bibnamefont {Xu}}, \bibinfo {author} {\bibfnamefont {G.~D.}\ \bibnamefont {Gu}},\ and\ \bibinfo {author} {\bibfnamefont {D.}~\bibnamefont {Colson}},\ }\href {https://doi.org/10.1103/PhysRevB.82.144516} {\bibfield  {journal} {\bibinfo  {journal} {Physical Review B}\ }\textbf {\bibinfo {volume} {82}},\ \bibinfo {pages} {144516} (\bibinfo {year} {2010})}\BibitemShut {NoStop}%
\bibitem [{\citenamefont {Saichu}\ \emph {et~al.}(2009)\citenamefont {Saichu}, \citenamefont {Mahns}, \citenamefont {Goos}, \citenamefont {Binder}, \citenamefont {May}, \citenamefont {Singer}, \citenamefont {Schulz}, \citenamefont {Rusydi}, \citenamefont {Unterhinninghofen}, \citenamefont {Manske}, \citenamefont {Guptasarma}, \citenamefont {Williamsen},\ and\ \citenamefont {Rübhausen}}]{saichu_two-component_2009}%
  \BibitemOpen
  \bibfield  {author} {\bibinfo {author} {\bibfnamefont {R.~P.}\ \bibnamefont {Saichu}}, \bibinfo {author} {\bibfnamefont {I.}~\bibnamefont {Mahns}}, \bibinfo {author} {\bibfnamefont {A.}~\bibnamefont {Goos}}, \bibinfo {author} {\bibfnamefont {S.}~\bibnamefont {Binder}}, \bibinfo {author} {\bibfnamefont {P.}~\bibnamefont {May}}, \bibinfo {author} {\bibfnamefont {S.~G.}\ \bibnamefont {Singer}}, \bibinfo {author} {\bibfnamefont {B.}~\bibnamefont {Schulz}}, \bibinfo {author} {\bibfnamefont {A.}~\bibnamefont {Rusydi}}, \bibinfo {author} {\bibfnamefont {J.}~\bibnamefont {Unterhinninghofen}}, \bibinfo {author} {\bibfnamefont {D.}~\bibnamefont {Manske}}, \bibinfo {author} {\bibfnamefont {P.}~\bibnamefont {Guptasarma}}, \bibinfo {author} {\bibfnamefont {M.~S.}\ \bibnamefont {Williamsen}},\ and\ \bibinfo {author} {\bibfnamefont {M.}~\bibnamefont {Rübhausen}},\ }\href {https://doi.org/10.1103/PhysRevLett.102.177004} {\bibfield  {journal} {\bibinfo  {journal} {Physical Review Letters}\ }\textbf {\bibinfo {volume}
  {102}},\ \bibinfo {pages} {177004} (\bibinfo {year} {2009})}\BibitemShut {NoStop}%
\bibitem [{\citenamefont {Rothwarf}\ and\ \citenamefont {Taylor}(1967)}]{rothwarf_measurement_1967}%
  \BibitemOpen
  \bibfield  {author} {\bibinfo {author} {\bibfnamefont {A.}~\bibnamefont {Rothwarf}}\ and\ \bibinfo {author} {\bibfnamefont {B.~N.}\ \bibnamefont {Taylor}},\ }\href {https://doi.org/10.1103/PhysRevLett.19.27} {\bibfield  {journal} {\bibinfo  {journal} {Physical Review Letters}\ }\textbf {\bibinfo {volume} {19}},\ \bibinfo {pages} {27} (\bibinfo {year} {1967})}\BibitemShut {NoStop}%
\bibitem [{\citenamefont {Kabanov}\ \emph {et~al.}(2005)\citenamefont {Kabanov}, \citenamefont {Demsar},\ and\ \citenamefont {Mihailovic}}]{kabanov_kinetics_2005}%
  \BibitemOpen
  \bibfield  {author} {\bibinfo {author} {\bibfnamefont {V.~V.}\ \bibnamefont {Kabanov}}, \bibinfo {author} {\bibfnamefont {J.}~\bibnamefont {Demsar}},\ and\ \bibinfo {author} {\bibfnamefont {D.}~\bibnamefont {Mihailovic}},\ }\href {https://doi.org/10.1103/PhysRevLett.95.147002} {\bibfield  {journal} {\bibinfo  {journal} {Physical Review Letters}\ }\textbf {\bibinfo {volume} {95}},\ \bibinfo {pages} {147002} (\bibinfo {year} {2005})}\BibitemShut {NoStop}%
\bibitem [{\citenamefont {Demsar}\ \emph {et~al.}(2003)\citenamefont {Demsar}, \citenamefont {Averitt}, \citenamefont {Taylor}, \citenamefont {Kabanov}, \citenamefont {Kang}, \citenamefont {Kim}, \citenamefont {Choi},\ and\ \citenamefont {Lee}}]{demsar_pair-breaking_2003}%
  \BibitemOpen
  \bibfield  {author} {\bibinfo {author} {\bibfnamefont {J.}~\bibnamefont {Demsar}}, \bibinfo {author} {\bibfnamefont {R.~D.}\ \bibnamefont {Averitt}}, \bibinfo {author} {\bibfnamefont {A.~J.}\ \bibnamefont {Taylor}}, \bibinfo {author} {\bibfnamefont {V.~V.}\ \bibnamefont {Kabanov}}, \bibinfo {author} {\bibfnamefont {W.~N.}\ \bibnamefont {Kang}}, \bibinfo {author} {\bibfnamefont {H.~J.}\ \bibnamefont {Kim}}, \bibinfo {author} {\bibfnamefont {E.~M.}\ \bibnamefont {Choi}},\ and\ \bibinfo {author} {\bibfnamefont {S.~I.}\ \bibnamefont {Lee}},\ }\href {https://doi.org/10.1103/PhysRevLett.91.267002} {\bibfield  {journal} {\bibinfo  {journal} {Physical Review Letters}\ }\textbf {\bibinfo {volume} {91}},\ \bibinfo {pages} {267002} (\bibinfo {year} {2003})}\BibitemShut {NoStop}%
\bibitem [{\citenamefont {{A. Shvaika}}\ \emph {et~al.}(2018)\citenamefont {{A. Shvaika}}, \citenamefont {{O. Matveev,}}, \citenamefont {{T. P. Devereaux}},\ and\ \citenamefont {{J. F. Freericks}}}]{shvaika_interpreting_2018}%
  \BibitemOpen
  \bibfield  {author} {\bibinfo {author} {\bibnamefont {{A. Shvaika}}}, \bibinfo {author} {\bibnamefont {{O. Matveev,}}}, \bibinfo {author} {\bibnamefont {{T. P. Devereaux}}},\ and\ \bibinfo {author} {\bibnamefont {{J. F. Freericks}}},\ }\href {https://doi.org/10.5488/CMP.21.33707} {\bibfield  {journal} {\bibinfo  {journal} {Condensed Matter Physics}\ }\textbf {\bibinfo {volume} {21}},\ \bibinfo {pages} {33707} (\bibinfo {year} {2018})}\BibitemShut {NoStop}%
\bibitem [{\citenamefont {Matveev}\ \emph {et~al.}(2019)\citenamefont {Matveev}, \citenamefont {Shvaika}, \citenamefont {Devereaux},\ and\ \citenamefont {Freericks}}]{matveev_stroboscopic_2019}%
  \BibitemOpen
  \bibfield  {author} {\bibinfo {author} {\bibfnamefont {O.}~\bibnamefont {Matveev}}, \bibinfo {author} {\bibfnamefont {A.}~\bibnamefont {Shvaika}}, \bibinfo {author} {\bibfnamefont {T.}~\bibnamefont {Devereaux}},\ and\ \bibinfo {author} {\bibfnamefont {J.}~\bibnamefont {Freericks}},\ }\href {https://doi.org/10.1103/PhysRevLett.122.247402} {\bibfield  {journal} {\bibinfo  {journal} {Physical Review Letters}\ }\textbf {\bibinfo {volume} {122}},\ \bibinfo {pages} {247402} (\bibinfo {year} {2019})}\BibitemShut {NoStop}%
\bibitem [{\citenamefont {Glier}\ \emph {et~al.}(2025)\citenamefont {Glier}, \citenamefont {Tian}, \citenamefont {Rerrer}, \citenamefont {Westphal}, \citenamefont {Lüllau}, \citenamefont {Feng}, \citenamefont {Dolgner}, \citenamefont {Haenel}, \citenamefont {Zonno}, \citenamefont {Eisaki}, \citenamefont {Greven}, \citenamefont {Damascelli}, \citenamefont {Kaiser}, \citenamefont {Manske},\ and\ \citenamefont {Rübhausen}}]{glier_non-equilibrium_2025}%
  \BibitemOpen
  \bibfield  {author} {\bibinfo {author} {\bibfnamefont {T.~E.}\ \bibnamefont {Glier}}, \bibinfo {author} {\bibfnamefont {S.}~\bibnamefont {Tian}}, \bibinfo {author} {\bibfnamefont {M.}~\bibnamefont {Rerrer}}, \bibinfo {author} {\bibfnamefont {L.}~\bibnamefont {Westphal}}, \bibinfo {author} {\bibfnamefont {G.}~\bibnamefont {Lüllau}}, \bibinfo {author} {\bibfnamefont {L.}~\bibnamefont {Feng}}, \bibinfo {author} {\bibfnamefont {J.}~\bibnamefont {Dolgner}}, \bibinfo {author} {\bibfnamefont {R.}~\bibnamefont {Haenel}}, \bibinfo {author} {\bibfnamefont {M.}~\bibnamefont {Zonno}}, \bibinfo {author} {\bibfnamefont {H.}~\bibnamefont {Eisaki}}, \bibinfo {author} {\bibfnamefont {M.}~\bibnamefont {Greven}}, \bibinfo {author} {\bibfnamefont {A.}~\bibnamefont {Damascelli}}, \bibinfo {author} {\bibfnamefont {S.}~\bibnamefont {Kaiser}}, \bibinfo {author} {\bibfnamefont {D.}~\bibnamefont {Manske}},\ and\ \bibinfo {author} {\bibfnamefont {M.}~\bibnamefont {Rübhausen}},\ }\href {https://doi.org/10.1038/s41467-025-62245-4}
  {\bibfield  {journal} {\bibinfo  {journal} {Nature Communications}\ }\textbf {\bibinfo {volume} {16}},\ \bibinfo {pages} {7027} (\bibinfo {year} {2025})}\BibitemShut {NoStop}%
\bibitem [{\citenamefont {Venturini}\ \emph {et~al.}(2002)\citenamefont {Venturini}, \citenamefont {Opel}, \citenamefont {Devereaux}, \citenamefont {Freericks}, \citenamefont {Tüttő}, \citenamefont {Revaz}, \citenamefont {Walker}, \citenamefont {Berger}, \citenamefont {Forró},\ and\ \citenamefont {Hackl}}]{venturini_observation_2002}%
  \BibitemOpen
  \bibfield  {author} {\bibinfo {author} {\bibfnamefont {F.}~\bibnamefont {Venturini}}, \bibinfo {author} {\bibfnamefont {M.}~\bibnamefont {Opel}}, \bibinfo {author} {\bibfnamefont {T.~P.}\ \bibnamefont {Devereaux}}, \bibinfo {author} {\bibfnamefont {J.~K.}\ \bibnamefont {Freericks}}, \bibinfo {author} {\bibfnamefont {I.}~\bibnamefont {Tüttő}}, \bibinfo {author} {\bibfnamefont {B.}~\bibnamefont {Revaz}}, \bibinfo {author} {\bibfnamefont {E.}~\bibnamefont {Walker}}, \bibinfo {author} {\bibfnamefont {H.}~\bibnamefont {Berger}}, \bibinfo {author} {\bibfnamefont {L.}~\bibnamefont {Forró}},\ and\ \bibinfo {author} {\bibfnamefont {R.}~\bibnamefont {Hackl}},\ }\href {https://doi.org/10.1103/PhysRevLett.89.107003} {\bibfield  {journal} {\bibinfo  {journal} {Physical Review Letters}\ }\textbf {\bibinfo {volume} {89}},\ \bibinfo {pages} {107003} (\bibinfo {year} {2002})}\BibitemShut {NoStop}%
\bibitem [{\citenamefont {Sacuto}\ \emph {et~al.}(2013)\citenamefont {Sacuto}, \citenamefont {Gallais}, \citenamefont {Cazayous}, \citenamefont {Méasson}, \citenamefont {Gu},\ and\ \citenamefont {Colson}}]{sacuto_new_2013}%
  \BibitemOpen
  \bibfield  {author} {\bibinfo {author} {\bibfnamefont {A.}~\bibnamefont {Sacuto}}, \bibinfo {author} {\bibfnamefont {Y.}~\bibnamefont {Gallais}}, \bibinfo {author} {\bibfnamefont {M.}~\bibnamefont {Cazayous}}, \bibinfo {author} {\bibfnamefont {M.-A.}\ \bibnamefont {Méasson}}, \bibinfo {author} {\bibfnamefont {G.~D.}\ \bibnamefont {Gu}},\ and\ \bibinfo {author} {\bibfnamefont {D.}~\bibnamefont {Colson}},\ }\href {https://doi.org/10.1088/0034-4885/76/2/022502} {\bibfield  {journal} {\bibinfo  {journal} {Reports on Progress in Physics}\ }\textbf {\bibinfo {volume} {76}},\ \bibinfo {pages} {022502} (\bibinfo {year} {2013})}\BibitemShut {NoStop}%
\bibitem [{\citenamefont {Nicol}\ and\ \citenamefont {Carbotte}(2003)}]{nicol_comparison_2003}%
  \BibitemOpen
  \bibfield  {author} {\bibinfo {author} {\bibfnamefont {E.~J.}\ \bibnamefont {Nicol}}\ and\ \bibinfo {author} {\bibfnamefont {J.~P.}\ \bibnamefont {Carbotte}},\ }\href {https://doi.org/10.1103/PhysRevB.67.214506} {\bibfield  {journal} {\bibinfo  {journal} {Physical Review B}\ }\textbf {\bibinfo {volume} {67}},\ \bibinfo {pages} {214506} (\bibinfo {year} {2003})}\BibitemShut {NoStop}%
\bibitem [{\citenamefont {Owen}\ and\ \citenamefont {Scalapino}(1972)}]{owen_superconducting_1972}%
  \BibitemOpen
  \bibfield  {author} {\bibinfo {author} {\bibfnamefont {C.~S.}\ \bibnamefont {Owen}}\ and\ \bibinfo {author} {\bibfnamefont {D.~J.}\ \bibnamefont {Scalapino}},\ }\href {https://doi.org/10.1103/PhysRevLett.28.1559} {\bibfield  {journal} {\bibinfo  {journal} {Physical Review Letters}\ }\textbf {\bibinfo {volume} {28}},\ \bibinfo {pages} {1559} (\bibinfo {year} {1972})}\BibitemShut {NoStop}%
\bibitem [{\citenamefont {Smallwood}\ \emph {et~al.}(2014)\citenamefont {Smallwood}, \citenamefont {Zhang}, \citenamefont {Miller}, \citenamefont {Jozwiak}, \citenamefont {Eisaki}, \citenamefont {Lee},\ and\ \citenamefont {Lanzara}}]{smallwood_time-_2014}%
  \BibitemOpen
  \bibfield  {author} {\bibinfo {author} {\bibfnamefont {C.~L.}\ \bibnamefont {Smallwood}}, \bibinfo {author} {\bibfnamefont {W.}~\bibnamefont {Zhang}}, \bibinfo {author} {\bibfnamefont {T.~L.}\ \bibnamefont {Miller}}, \bibinfo {author} {\bibfnamefont {C.}~\bibnamefont {Jozwiak}}, \bibinfo {author} {\bibfnamefont {H.}~\bibnamefont {Eisaki}}, \bibinfo {author} {\bibfnamefont {D.-H.}\ \bibnamefont {Lee}},\ and\ \bibinfo {author} {\bibfnamefont {A.}~\bibnamefont {Lanzara}},\ }\href {https://doi.org/10.1103/PhysRevB.89.115126} {\bibfield  {journal} {\bibinfo  {journal} {Physical Review B}\ }\textbf {\bibinfo {volume} {89}},\ \bibinfo {pages} {115126} (\bibinfo {year} {2014})}\BibitemShut {NoStop}%
\bibitem [{\citenamefont {Gedik}\ \emph {et~al.}(2003)\citenamefont {Gedik}, \citenamefont {Orenstein}, \citenamefont {Liang}, \citenamefont {Bonn},\ and\ \citenamefont {Hardy}}]{gedik_diffusion_2003}%
  \BibitemOpen
  \bibfield  {author} {\bibinfo {author} {\bibfnamefont {N.}~\bibnamefont {Gedik}}, \bibinfo {author} {\bibfnamefont {J.}~\bibnamefont {Orenstein}}, \bibinfo {author} {\bibfnamefont {R.}~\bibnamefont {Liang}}, \bibinfo {author} {\bibfnamefont {D.~A.}\ \bibnamefont {Bonn}},\ and\ \bibinfo {author} {\bibfnamefont {W.~N.}\ \bibnamefont {Hardy}},\ }\href {https://doi.org/10.1126/science.1083038} {\bibfield  {journal} {\bibinfo  {journal} {Science}\ }\textbf {\bibinfo {volume} {300}},\ \bibinfo {pages} {1410} (\bibinfo {year} {2003})}\BibitemShut {NoStop}%
\bibitem [{\citenamefont {Valenzuela}\ and\ \citenamefont {Bascones}(2007)}]{valenzuela_phenomenological_2007}%
  \BibitemOpen
  \bibfield  {author} {\bibinfo {author} {\bibfnamefont {B.}~\bibnamefont {Valenzuela}}\ and\ \bibinfo {author} {\bibfnamefont {E.}~\bibnamefont {Bascones}},\ }\href {https://doi.org/10.1103/PhysRevLett.98.227002} {\bibfield  {journal} {\bibinfo  {journal} {Physical Review Letters}\ }\textbf {\bibinfo {volume} {98}},\ \bibinfo {pages} {227002} (\bibinfo {year} {2007})}\BibitemShut {NoStop}%
\end{thebibliography}
\end{document}